# Early Planet Formation in Embedded Disks (eDisk). VIII. A Small Protostellar Disk around the Extremely Low-Mass and Young Class 0 Protostar, IRAS 15398-3359

Travis J. Thieme,[1,2,3] Shih-Ping Lai,[1,2,3,4] Nagayoshi Ohashi,[4] John J. Tobin,[5] Jes K. Jørgensen,[6] Jinshi Sai (Insa Choi),[4] Yusuke Aso,[7] Jonathan P. Williams,[8] Yoshihide Yamato,[9] Yuri Aikawa,[9] Itziar de Gregorio-Monsalvo,[10] Ilseung Han,[11,7] Woojin Kwon,[12,13] Chang Won Lee,[11,7] Jeong-Eun Lee,[14] Zhi-Yun Li,[15] Zhe-Yu Daniel Lin,[15] Leslie W. Looney,[16] Suchitra Narayanan,[8] Nguyen Thi Phuong,[7,17] Adele L. Plunkett,[5] Alejandro Santamaría-Miranda,[10] Rajeeb Sharma,[6] Shigehisa Takakuwa,[18,4] and Hsi-Wei Yen[4]

[1] Institute of Astronomy, National Tsing Hua University, No. 101, Section 2, Kuang-Fu Road, Hsinchu 30013, Taiwan
[2] Center for Informatics and Computation in Astronomy, National Tsing Hua University, No. 101, Section 2, Kuang-Fu Road, Hsinchu 30013, Taiwan
[3] Department of Physics, National Tsing Hua University, No. 101, Section 2, Kuang-Fu Road, Hsinchu 30013, Taiwan
[4] Academia Sinica Institute of Astronomy and Astrophysics, 11F of Astro-Math Bldg, No. 1, Sec. 4, Roosevelt Rd, Taipei 10617, Taiwan
[5] National Radio Astronomy Observatory, 520 Edgemont Rd., Charlottesville, VA 22903, USA
[6] Niels Bohr Institute, University of Copenhagen, Øster Voldgade 5-7, 1350, Copenhagen, Denmark
[7] Korea Astronomy and Space Science Institute, 776 Daedeok-daero, Yuseong-gu, Daejeon 34055, Republic of Korea
[8] Institute for Astronomy, University of Hawaiʻi at Mānoa, 2680 Woodlawn Dr., Honolulu, HI 96822, USA
[9] Department of Astronomy, Graduate School of Science, The University of Tokyo, 7-3-1 Hongo, Bunkyo-ku, Tokyo 113-0033, Japan
[10] European Southern Observatory, Alonso de Cordova 3107, Casilla 19, Vitacura, Santiago, Chile
[11] Division of Astronomy and Space Science, University of Science and Technology, 217 Gajeong-ro, Yuseong-gu, Daejeon 34113, Republic of Korea
[12] Department of Earth Science Education, Seoul National University, 1 Gwanak-ro, Gwanak-gu, Seoul 08826, Republic of Korea
[13] SNU Astronomy Research Center, Seoul National University, 1 Gwanak-ro, Gwanak-gu, Seoul 08826, Republic of Korea
[14] Department of Physics and Astronomy, Seoul National University, 1 Gwanak-ro, Gwanak-gu, Seoul 08826, Republic of Korea
[15] University of Virginia, 530 McCormick Rd., Charlottesville, Virginia 22904, USA
[16] Department of Astronomy, University of Illinois, 1002 West Green St, Urbana, IL 61801, USA
[17] Department of Astrophysics, Vietnam National Space Center, Vietnam Academy of Science and Techonology, 18 Hoang Quoc Viet, Cau Giay, Hanoi, Vietnam
[18] Department of Physics and Astronomy, Graduate School of Science and Engineering, Kagoshima University, 1-21-35 Korimoto, Kagoshima,Kagoshima 890-0065, Japan

## ABSTRACT

Protostellar disks are an ubiquitous part of the star formation process and the future sites of planet formation. As part of the Early Planet Formation in Embedded Disks (eDisk) large program, we present high-angular resolution dust continuum ($\sim 40$ mas) and molecular line ($\sim 150$ mas) observations of the Class 0 protostar, IRAS 15398-3359. The dust continuum is small, compact, and centrally peaked, while more extended dust structures are found in the outflow directions. We perform a 2D Gaussian fitting to find the deconvolved size and $2\sigma$ radius of the dust disk to be $4.5 \times 2.8$ au and 3.8 au, respectively. We estimate the gas+dust disk mass assuming optically thin continuum emission to be $0.6 - 1.8\,M_{\rm jup}$, indicating a very low-mass disk. The CO isotopologues trace components of the outflows and inner envelope, while SO traces a compact, rotating disk-like component. Using several rotation curve fittings on the PV diagram of the SO emission, the lower limits of the protostellar mass and gas disk radius are $0.022\,M_\odot$ and 31.2 au from our Modified 2 single power-law fitting. A conservative upper limit of the protostellar mass is inferred to be $0.1\,M_\odot$. The protostellar mass-accretion rate and the specific angular momentum at the protostellar disk edge are found to be between $1.3 - 6.1 \times 10^{-6}\,M_\odot\,{\rm yr}^{-1}$ and $1.2 - 3.8 \times 10^{-4}\,{\rm km\,s^{-1}\,pc}$, respectively, with an age estimated between $0.4 - 7.5 \times 10^4$ yr. At this young age with no clear substructures in the disk, planet formation would likely not yet have started. This study highlights the importance of high-resolution observations and systematic fitting procedures when deriving dynamical properties of deeply embedded Class 0 protostars.

## 1. INTRODUCTION

Protostellar disks are essential for transferring mass to the central protostar and forming future planetary systems. Kinematic observations of these disks provide a way to determine dynamical protostellar masses, which is a key aspect in understanding the evolution of young protostars. T Tauri and pre-main sequence (Class II/III) protostars harbor clear, rotationally-supported disks, allowing for derivations of their dynamical masses (e.g., Simon et al. 2000; Guilloteau et al. 2014; Sheehan et al. 2019; Boyden & Eisner 2020). For deeply embedded (Class 0/I) protostars, dynamical mass derivations are complicated by the kinematics of the larger scale envelopes (e.g., Yen et al. 2015a). Nonetheless, interferometric observations have provided a possibility to detect rotationally-supported Keplerian disks at these early stages (e.g., Tobin et al. 2012; Murillo et al. 2013; Yen et al. 2014; Ohashi et al. 2014; Aso et al. 2015; Yen et al. 2017a,b, 2019; Maret et al. 2020; Sai et al. 2020). Thus, executing high-angular resolution observations can shed light on the formation and early evolution of protostellar disks.

Rotationally-supported disks in the Class 0 stage are typically studied via position-velocity (PV) diagrams of optically-thin molecular line emission along the inferred disk axis. The PV structure is then compared to and/or fit with a Keplerian rotation power-law profile (e.g., Ohashi et al. 2014; Aso et al. 2017; Yen et al. 2017a; Maret et al. 2020). A rotational profile proportional to $v \sim r^{-1/2}$ corresponds to a Keplerian rotating disk, whereas a rotational profile proportional to $v \sim r^{-1}$ indicates infalling material with conserved angular momentum. The methodology for fitting PV diagrams has been developed over the years and tested against simulations (Harsono et al. 2015; Aso & Machida 2020). These results confirm that fitting peak positions/velocities in the PV diagram can accurately recover crucial properties, in particular, the dynamical protostellar mass.

Additionally, deeply embedded disks likely also represent the places where planet formation is initiated. Gaps and rings are clearly revealed by observations of later-stage disks in both dust continuum (e.g., ALMA Partnership et al. 2015; Andrews et al. 2018) and molecular line emission (e.g., Öberg et al. 2021). These substructures are thought to be where planet formation is currently ongoing. Recent observations of protostars that are still in the embedded stages have also revealed the same types of substructures (Segura-Cox et al. 2020; Sheehan et al. 2020), indicating that planet formation may begin earlier than previously thought. These results are substantiated by studies indicating early grain-growth (Harsono et al. 2018) and that the dust reservoir in these early disks is enough to produce observed exoplanetary systems (Tychoniec et al. 2020). However, these ring and gap structures can also form due from gravitational instabilities, tidal processes, dust processes and more (see Bae et al. 2022 for a full review).

In this paper, we present new, high-angular resolution Atacama Large Millimeter/submillimeter Array (ALMA) observations of the young protostar, IRAS 15398-3359, as part of the Early Planet Formation in Embedded Disk (eDisk) large program. IRAS 15398-3359 (hereafter, IRAS15398), is located in the B228 region of the Lupus I molecular cloud. Several recent studies derive the distance to Lupus I using *Gaia* DR2 data (Dzib et al. 2018; Zucker et al. 2020; Galli et al. 2020; Santamaría-Miranda et al. 2021). We adopt a distance of 155.5 pc by taking the average distances of sources and positions closest to the Right Ascension (RA) and Declination (Dec) of IRAS15398 from these previous studies. Based on recent SED fittings, it is classified as a Class 0 protostar with $T_{\rm bol} = 50\,{\rm K}$ and $L_{\rm bol} = 1.4\,L_\odot$ (Ohashi et al. 2023). This is consistent with classifications in previous studies, which have also considered other evolutionary indicators, such as the chemistry and outflow properties (Bjerkeli et al. 2016a; Vazzano et al. 2021).

Kinematic studies of IRAS15398 analyze the infalling-rotating envelope around an inferred Keplerian disk (Oya et al. 2014; Yen et al. 2017a; Okoda et al. 2018). Initial studies were not able to resolve a Keplerian disk, and instead infer an upper limit on the protostellar mass using simple envelope models (Oya et al. 2014; Yen et al. 2017a). Oya et al. (2014) use an infalling-rotating envelope model to constrain the central protostellar mass to $M_\star \leq 0.09\,M_\odot$ using observations of several $H_2CO$ transitions with an angular resolution of $0.60'' \times 0.44''$. Additionally, Yen et al. (2017a) use $C^{18}O$ ($J = 2 \rightarrow 1$) emission at an angular resolution of $0.53'' \times 0.49''$ to fit a rotational power-law profile to points from their PV diagram. Although they found a power-law index of $-1.0 \pm 0.06$, indicating the emission traces the infalling protostellar envelope, the disk was

**Table 1.** Summary of eDisk Observations of IRAS 15398-3359

| **Parameters** | **Short Baseline (SB)** | **Long Baseline (LB)** |
|---|---|---|
| Project I.D. | 2019.A.00034.S | 2019.1.00261.L |
| Observation Dates | 2021 May 06 | 2021 Aug 09, 2021 Aug 13-14, 2021 Oct 20 |
| Array Configuration | C43-5 | C43-8 |
| Number of Antennas | 45 | 39, 42, 43 |
| Min - Max Baseline (m) | 15-2517 | 70-8282, 70-8282, 46-8983 |
| Phase Center (ICRS) | $15^{h}43^{m}02.^{s}24$ $-34^{\circ}09'06.''8$ | |
| | **Calibrators** | |
| Bandpass | J1337-1257 | J1337-1257, J1517-2422, J1427-4206 |
| Flux | J1337-1257 | J1337-1257, J1517-2422, J1427-4206 |
| Phase | J1610-3958 | J1534-3526 |
| Pointing | J1337-1257, J1610-3958 | J1337-1257, J1457-3539, J1517-2422, J1427-4206 |

Note—See Ohashi et al. (2023) for a detailed summary of the eDisk program, observations and data reduction procedure.

not resolved in their observations. An upper limit on the protostellar mass was found to be $\sim 0.01\,M_{\odot}$, along with an upper limit disk radius of $20^{+50}_{-20}$ au and an inclination angle of 70°. However, if the infall is slower than the free-fall assumption used in their analysis, the protostellar mass would be a lower limit. Higher resolution (0.2″) observations were later performed and a rotationally-supported disk was inferred by comparing the PV emission structure of SO ($J_N = 7_6 \rightarrow 6_5$) to a Keplerian rotation power-law profile (Okoda et al. 2018). A profile assuming a dynamical protostellar mass of $0.007^{+0.004}_{-0.003}\,M_{\odot}$ and an inclination angle of 70° matched well by eye with the overall PV diagram emission. These studies indicate a relatively small, difficult to detect Keplerian disk around an extremely low-mass protostar. Systematic fitting procedures of the protostellar disk emission in the PV diagram at a higher resolution have yet to be implemented for this source to derive its properties.

A number of different studies on IRAS15398 provide clues to the interesting nature of the disk and protostar. The core-scale magnetic field strength has been derived to be 78 $\mu$G using polarization observations from the SOFIA telescope (Redaelli et al. 2019). Using the ratio of the turbulent and uniform magnetic field components, they conclude the core is strongly magnetized. In this case, magnetic braking could be efficient at removing the angular momentum of infalling material and thus hindering the growth of a large disk (e.g., Mellon & Li 2008a; Hennebelle & Ciardi 2009). The current primary outflow is estimated to have a dynamical timescale of $\lesssim$ 1000 yr (Bjerkeli et al. 2016a), and also is episodic in nature (Vazzano et al. 2021). This is complimented by evidence of a past episodic accretion event in the last 100-1000 yr (Jørgensen et al. 2013; Bjerkeli et al. 2016b). These previous studies further highlight the importance of our new higher-angular resolution observations to understand the nature of the IRAS15398 system.

This paper is organized as follows. Section 2 describes the eDisk observations of IRAS15398 and a summary of the data reduction process. Section 3 presents the dust continuum and molecular line maps. We analyze the nature of the dust continuum and the dynamical properties of the protostellar disk in Section 4. In Section 5 we discuss the implications of our results and compare to the previous studies of IRAS15398. Finally, we conclude our results in Section 6.

## 2. OBSERVATIONS & DATA REDUCTION

Observations of IRAS15398 were carried out with ALMA as part of the eDisk large program (project ID: 2019.1.00261.L; PI: N. Ohashi) with additional short baseline data taken as part of a complementary Director Discretionary Time (DDT) program (project ID: 2019.A.00034.S). The general characteristics of the observations are summarized in Table 1. An extensive description of the procedure for the eDisk data reduction, including links to the scripts, are provided in Ohashi et al. (2023). In this section, we summarize the key aspects of the data reduction and imaging for IRAS15398.

Long baseline observations were taken in three execution blocks on August 9th, 2021 with 39 antennas, August 13th-14th, 2021 with 42 antennas and October 20th, 2021 with 43 antennas. Additional short baseline observations were taken on May 6th, 2021 with 45 antennas. The data were reduced using the Common Astronomy Software Applications (CASA) package v6.2.1 (McMullin et al. 2007; CASA Team et al.

**Table 2.** Summary of SB+LB Dust Continuum and Molecular Line Maps

| Continuum / Line | Frequency (GHz) | Robust | Beamsize | Velocity Resolution ($km\,s^{-1}$) | Peak Intensity ($mJy\,beam^{-1}$) | RMS Noise ($mJy\,beam^{-1}$) |
|---|---|---|---|---|---|---|
| 1.3 mm continuum | 225 | −0.5 | $0\farcs043 \times 0\farcs036$ (−55.8°) | - | 5.95 | 0.04 |
| 1.3 mm continuum | 225 | +0.5 | $0\farcs084 \times 0\farcs071$ (+78.0°) | - | 7.10 | 0.02 |
| $C^{18}O$ ($J = 2 \rightarrow 1$) | 219.56035 | +1.5 | $0\farcs154 \times 0\farcs141$ (−82.6°) | 0.167 | 16.2 | 1.85 |
| SO ($J_N = 6_5 \rightarrow 5_4$) | 219.94944 | +1.5 | $0\farcs155 \times 0\farcs139$ (−84.6°) | 0.167 | 19.2 | 2.19 |
| $^{13}CO$ ($J = 2 \rightarrow 1$) | 220.39868 | +1.5 | $0\farcs146 \times 0\farcs132$ (−81.1°) | 0.167 | 31.4 | 2.33 |
| $^{12}CO$ ($J = 2 \rightarrow 1$) | 230.53800 | +1.5 | $0\farcs166 \times 0\farcs150$ (−81.4°) | 0.635 | 92.3 | 1.01 |

Note—The values listed here were obtained from the maps using short baseline + long baseline (SB+LB) observations. Dust continuum maps for all the robust values produced by the eDisk imaging script are shown in Appendix A. Short baseline-only (SB-only) observation maps of the molecular lines in this table are described and shown in Appendix C. The peak intensity of the continuum maps were found by using a 2″circular aperture around the center position, while for the line maps the same aperture was used but on the channel with the highest intensity. The rms noise of the continuum maps were found by using a 10″circular aperture in an emission-free area, while for the line maps the same aperture was used but around the center position on a line-free channel.

2022). Standardized eDisk reduction scripts were used for self-calibrating and imaging the short+long baseline (SB+LB) data and the short baseline only (SB-only)[1]. In both cases, the continuum data is self-calibrated first, followed by the molecular line data since the line self-calibration uses the self-calibration solutions from the continuum. In the case of SB-only data, five iterations of phase-only self-calibration were performed. For the combined SB+LB data, six iterations of phase-only self-calibration were performed on the SB-only visibilities, followed by four iterations of phase-only self-calibration on the combined SB+LB visibilities.

Continuum images were then produced using the CASA *tclean* task with Briggs weighting. Robust values of -2.0, -1.0, -0.5, 0.0, 0.5, 1.0, and 2.0 were used to find the image with the best balance between angular resolution and noise. Additional images with UV tapers of 3000k$\lambda$, 2000k$\lambda$, and 1000k$\lambda$ are also automatically produced by the script for robust values of 1.0 and 2.0. The molecular line visibilities were also aligned, scaled by flux between observations, as well as, continuum subtracted using the line-free channels. The self-calibration solutions from the continuum visibilities were then applied to the spectral lines and then cleaned using the CASA *tclean* task with Briggs weighting and a robust value of 1.5 to obtain better signal-to-noise. A summary of the representative maps that will be analyzed in this paper are listed in Table 2.

## 3. RESULTS

### 3.1. *Dust Continuum*

The high-angular resolution (SB+LB) 1.3 mm dust continuum maps for two robust values of +0.5 and -0.5 are shown in Figure 1. A gallery of the SB+LB continuum maps for each robust value produced by the imaging script are shown in Appendix A. The robust=+0.5 map shows extended dust emission out to a few 100 au, and has a peak intensity ($I_\nu^{\rm peak}$) and brightness temperature ($T_b^{\rm peak}$) of 7.10 mJy beam$^{-1}$ and 28.6 K, respectively. This extended emission is also seen in previous studies of this source (Oya et al. 2014; Yen et al. 2017a; Okoda et al. 2018), though, it is more resolved out in our image due to the higher angular resolution of $\sim$ 80 mas (12.4 au at 155.5 pc). The extended emission follows the overall direction of the outflow, which is indicated by the red and blue arrows in Figure 1 (left). The robust=-0.5 map shows a single, compact dust structure with an $I_\nu^{\rm peak}$ and $T_b^{\rm peak}$ of 5.95 mJy beam$^{-1}$ and 93.4 K, respectively. The angular resolution of $\sim$ 40 mas (6.2 au at 155.5 pc) gives us a resolved image with respect to the beam. This resolution is $\sim$10 times better resolution than the previous Band 6 observations reported in Yen et al. (2017a), and $\sim$4 times better than the most recent Band 6 observations reported by Okoda et al. (2021) and Okoda et al. (2023). There are four, low signal-to-noise structures at 3-5$\sigma$ extending out $\sim$2 au from the central continuum emission. These structures are contiguous with the larger scale dust emission seen in the robust=+0.5 map. In the gallery of continuum images (Figure A.1), we see that as we decrease the robust value in our image cleaning, these structures, along with the large scale emission, disappear from the image. This indicates the structures are related to the larger scale dust emission and just being resolved out as we increase the image spatial resolution.

[1] https://github.com/jjtobin/edisk

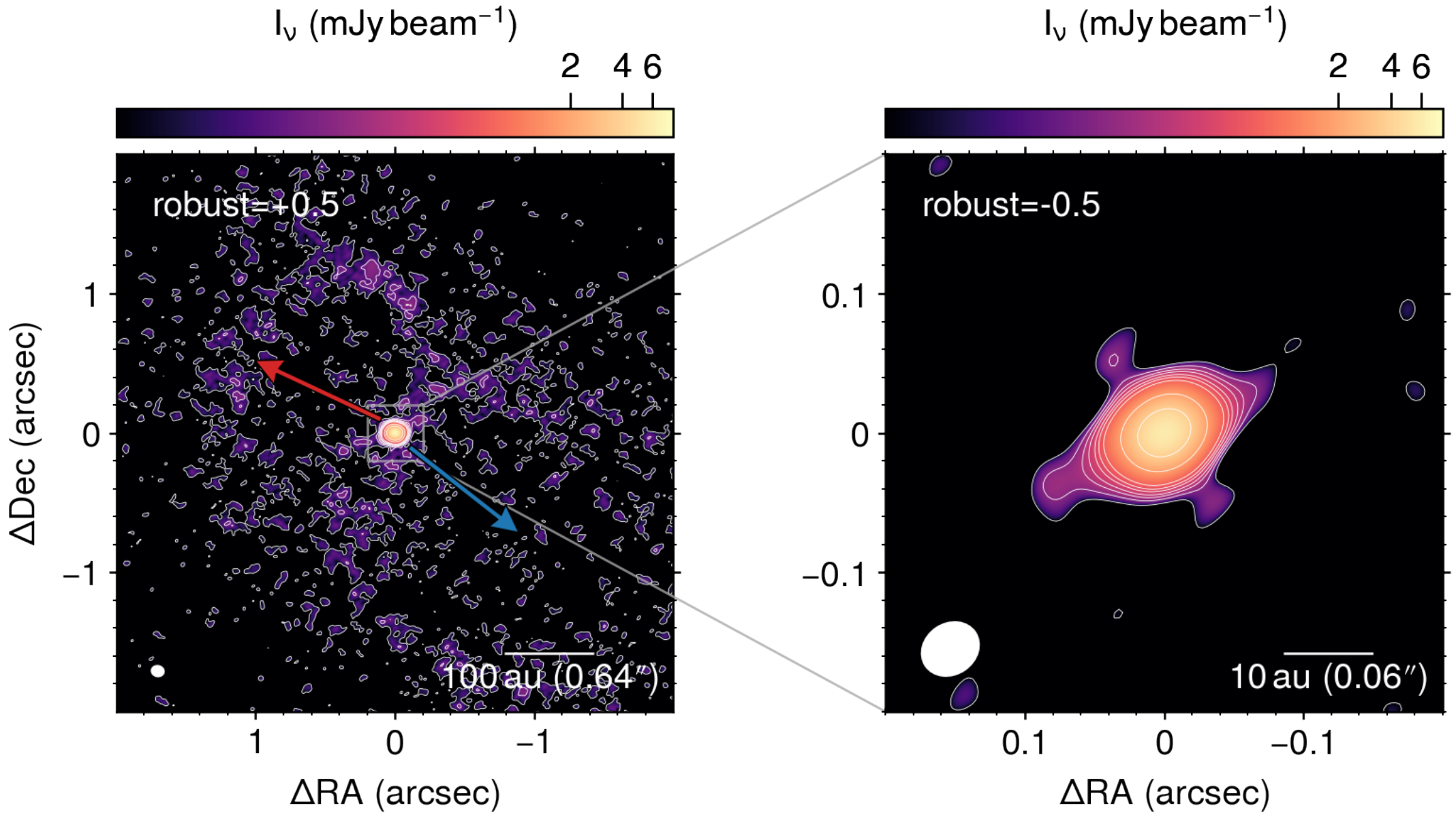

**Figure 1.** Representative 1.3 mm dust continuum maps of IRAS15398 with asinh (a=0.001) scaling. **(Left)** The robust=0.5 map showing an overall view of the extended dust emission. The blue and red arrows represent the overall blue and red-shifted outflow directions found by Vazzano et al. (2021), respectively. **(Right)** The robust=-0.5 map showing a zoomed-in view of the central compact structure. The contours in both maps are shown in steps 3, 5, 7, 9, 12, 15, 20, 50 and 100$\sigma_{\rm rms}$, where $\sigma_{\rm rms}$ is the rms noise of each map.

We measured the geometric parameters and the integrated flux density of the robust=-0.5 dust continuum map by performing a 2D Gaussian fit in the image plane using the CASA *imfit* task. The results are shown in Figure 2 and Table 3. The residual image in Figure 2 shows that the central emission is well represented by a 2D Gaussian, and there are no clear leftover substructures within the disk. We also checked this by making intensity cuts and radial profiles, but failed to see any sign of substructure in the disk. There are four leftover extended structures in the residual image corresponding to the extended continuum emission in the observation map. The coordinates at the peak intensity are $\alpha$(ICRS)=$15^{\rm h}43^{\rm m}02\overset{\rm s}{.}232$, $\delta$(ICRS)=-$34\overset{\circ}{.}09'\!.06''\!.96$. This position is adopted as the protostellar position of IRAS 15398-3359 for the figures and analysis in this paper. The deconvolved size ($\theta_{\rm maj} \times \theta_{\rm min}$) was found to be $29 \times 18$ mas ($4.5 \times 2.8$ au at a distance of 155.5 pc), with the position angle (north to east) of the major axis being 117.1°. By assuming a geometrically thin disk, the inclination angle can roughly be estimated by

$$i = \cos^{-1}\left(\theta_{\rm min}/\theta_{\rm maj}\right), \tag{1}$$

which gives an angle of $\sim$ 50.7°. The peak intensity and integrated flux density measured by the fitting are 5.86 mJy beam$^{-1}$ and 7.89 mJy, respectively.

The (gas+dust) disk mass can be estimated under the assumptions of entirely thermal and optically thin dust emission via

$$M_{\rm disk} = \frac{D^2 S_\nu}{\kappa_\nu B_\nu(T_{\rm dust})}, \tag{2}$$

where $D$ is the distance to the source, $S_\nu$ is the integrated flux density, $\kappa_\nu$ is the dust opacity adjusted for the gas-to-dust ratio, and $B_\nu$ is the Planck function for the dust temperature, $T_{\rm dust}$. We use the integrated flux density of 7.89 mJy from the Gaussian fitting and a distance of 155.5 pc. We adopt a gas-to-dust ratio of 100 for a 1.3 mm dust opacity value of $\kappa_\nu = 0.023\,{\rm cm^2\,g^{-1}}$ (Beckwith et al. 1990). For the dust temperature, we consider two cases. In the first, we use a $T_{\rm dust}$ of 20 K, which is a typical value adopted in studies of low-mass star formation (e.g., Jørgensen et al. 2009; Tobin et al. 2015). In order to give a more direct comparison of disk masses between sources in our eDisk sample, which have a wide range of protostellar luminosities, we also adopt a temperature distribution that scales with the bolometric luminosity ($L_{\rm bol}$) of the source. We use the following relation of

$$T_{\rm dust} = 43\,{\rm K}\left(\frac{L}{L_\odot}\right)^{0.25}, \tag{3}$$

which was derived using a grid of radiative transfer models (Tobin et al. 2020). For the measured $L_{\rm bol}$ of 1.4 $L_\odot$ (Ohashi et al. 2023), the resulting dust temperature is calculated to be 47 K. We calculate a (gas+dust) disk mass of $1.7 \times 10^{-3}\,M_\odot$ (1.8 $M_{\rm jup}$) for $T_{\rm dust} = 20$ K, and $6.1 \times 10^{-4}\,M_\odot$ (0.6 $M_{\rm jup}$) for $T_{\rm dust} =$

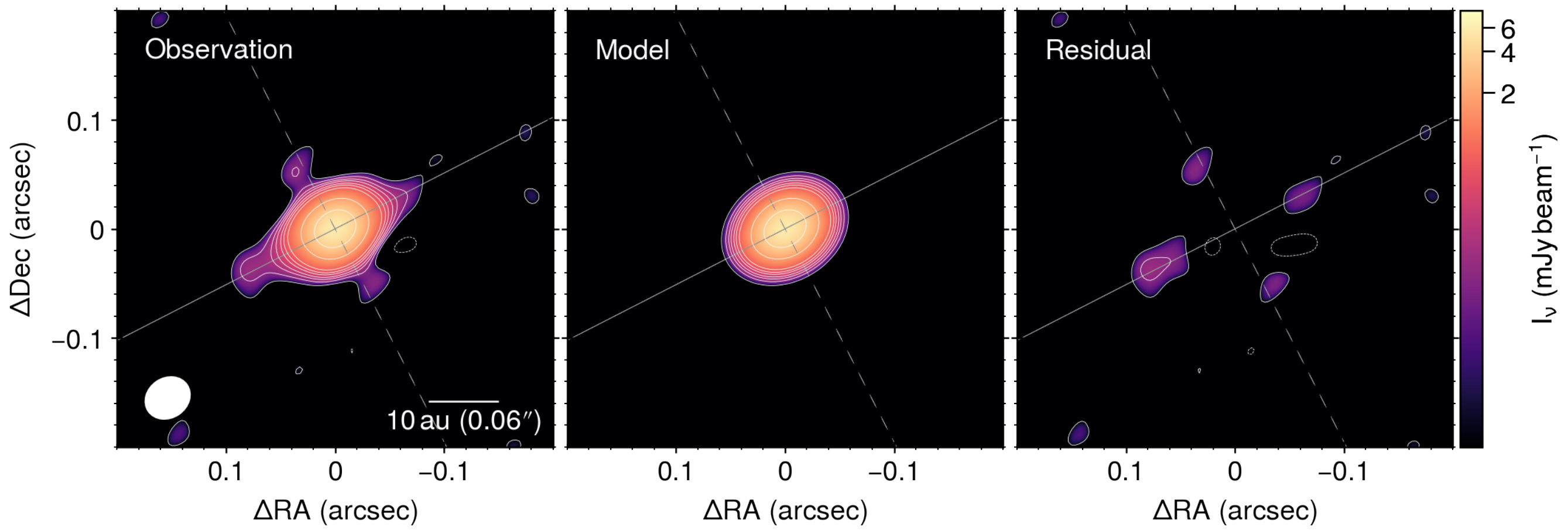

**Figure 2.** 2D Gaussian fitting results of IRAS 15398-3359. **(Left)** The robust=-0.5 map as shown in Figure 1. **(Center)** The 2D Gaussian fitting model results. The solid and dashed gray lines indicate the deconvolved position angle of the continuum major and minor axes, respectively. **(Right)** The residual map made by subtracting the model from the observations. The contours are shown in steps 3, 5, 7, 9, 12, 15, 20, 50 and 100$\sigma_{\rm rms}$, where $\sigma_{\rm rms}$ is the rms noise in the robust=-0.5 map.

47 K. If the dust is optically thick, these values would be lower limits.

**Table 3.** 2D Gaussian Fitting and Disk Mass

| Parameter | Value |
|---|---|
| **Peak Position** | |
| Right Ascension (ICRS) | $15^{\rm h}43^{\rm m}02\overset{\rm s}{.}232$ |
| Declination (ICRS) | $-34\overset{\circ}{.}09\overset{\prime}{.}06\overset{\prime\prime}{.}96$ |
| **Deconvolved Size** | |
| $\theta_{\rm maj}$ (mas) | $28.5 \pm 0.7$ |
| $\theta_{\rm min}$ (mas) | $18.1 \pm 0.6$ |
| $\theta_{\rm pa}$ (°) | $117.1 \pm 2.7$ |
| $i$ (°) | $50.7 \pm 2.0$ |
| **Peak Intensity & Integrated Flux Density** | |
| $I_\nu^{\rm fit}$ (mJy beam$^{-1}$) | $5.86 \pm 0.04$ |
| $S_\nu^{\rm fit}$ (mJy) | $7.89 \pm 0.09$ |
| **(Gas + Dust) Disk Mass** | |
| $M_{\rm disk}$ ($M_\odot$) | |
| - for $T_{\rm dust} = 20$ K | $1.69 \times 10^{-3}$ |
| - for $T_{\rm dust} = 47$ K | $6.14 \times 10^{-4}$ |

Note—These results were obtained using the robust=-0.5 dust continuum map.

### 3.2. *Molecular Lines*

The high resolution (SB+LB) $^{12}$CO (2−1), C$^{18}$O (2−1), $^{13}$CO (2 − 1), and SO ($6_5 - 5_4$) integrated-intensity and intensity-weighted velocity maps of IRAS15398 are shown in Figures 3 and 4, respectively. Both the integrated-intensity and intensity-weighted velocity maps are made using the *bettermoments* package (Teague & Foreman-Mackey 2018; Teague 2019). The centroid velocity is chosen to be 5.2 km s$^{-1}$ for the intensity-weighted velocity maps (Yen et al. 2017a). The emission above 3$\sigma$ in the datacubes were masked, along with the emission above 3$\sigma$ in the integrated-intensity maps. The angular resolution of $\sim$ 150 mas (23 au at 155.5 pc) is ~3 times better resolution than the previous Band 6 observations reported in Yen et al. (2017a), and at least ~1.3 times better than the more recent Band 6 observations reported by Okoda et al. (2018), Okoda et al. (2021) and Okoda et al. (2023). The estimated rms of the $^{12}$CO, C$^{18}$O, $^{13}$CO, and SO integrated-intensity maps are 4.5, 1.2, 2.3 and 1.9 mJy beam$^{-1}$ km s$^{-1}$, respectively. Channel maps of the SB+LB molecular line data are shown in Appendix B with their prominent features described in this section. SB-only maps of the aforementioned molecules are shown in Appendix C and are also briefly described in this section in comparison to their respective SB+LB maps. Several other molecular lines are also targeted by the eDisk program, including three transitions of H$_2$CO ($3_{0,3} - 2_{0,2}$, $3_{2,2} - 2_{2,1}$, $3_{2,1} - 2_{2,0}$), four transitions of c-C$_3$H$_2$ ($6_{0,6} - 5_{1,5}$, $6_{1,6} - 5_{0,5}$, $5_{1,4} - 4_{2,3}$, $5_{2,4} - 4_{1,3}$), and single transitions of CH$_3$OH ($4_2 - 3_1$ $E$), DCN (3 − 2) and SiO (5 − 4). In this paper, we focus on the highest spectral resolution CO and SO isotopologue lines.

#### 3.2.1. *$^{12}$CO (2–1)*

The large-scale $^{12}$CO emission is shown in Figures 3a & 4a. The outflow is clearly traced by $^{12}$CO and exhibits a slight bend downward on the northeastern red-shifted side and upward on the southwestern blue-shifted side. This is consistent with the precessing outflow previously interpreted by Vazzano et al. (2021). Additionally, it is visually clumpy in the integrated-intensity map. There is enhanced emission in the outer cavities near the source, especially on the upper side of the red-shifted outflow and on the lower side of the blue-shifted outflow (Figure 3a). There is a slightly blue-shifted patch of emission to the southeast at $\sim$ 1000 au and northwest at $\sim$ 700 au (Figure 4a). In the $^{12}$CO channel map, the south-east patch is seen in between 3.50 km s$^{-1}$ to 4.77 km s$^{-1}$, and seems to increase in velocity as the distance increases from the source (Figure B.1). The northwest patch appears in many of the blue-shifted channels and is aligned parallel to the red-shifted outflow. The red-shifted outflow shows a mixture of red and blue-shifted emission, hinting that the outflow is more likely in the plane-of-sky. The outflows are offset from the direction perpendicular to the disk. The position angles of the red and blue-shifted outflows were previously found to be $64.9 \pm 0.2°$ and $232.0 \pm 0.2°$, respectively (Vazzano et al. 2021). This leads to offsets of $117.1° - 64.9° = 52.2°$ for the red-shifted outflow and $232.0° - 117.1° = 114.9°$ for the blue-shifted outflow from the inferred position angle of the disk, where one would generally expect a $\sim$ 90° (perpendicular) offset. This offset could further indicate a possible precessing outflow caused by tidal interactions of a binary companion or asymmetric infall re-orienting the angular momentum axis of the disk (e.g., Lee 2020). Compared to the SB-only map (Figure C.1a & C.2a) the overall structure is the same. There are clumpy portions, and the extended structure to the south-east is also present.

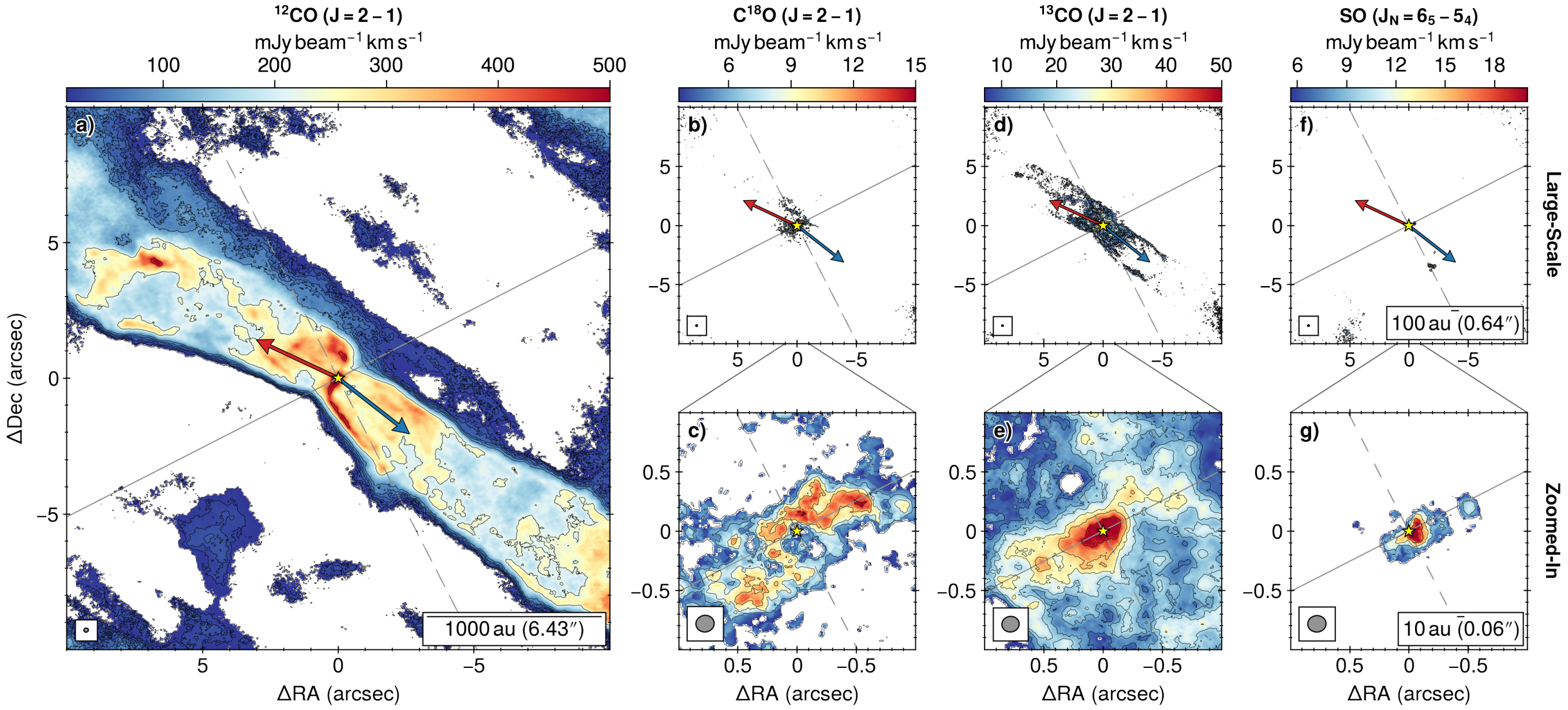

**Figure 3.** Integrated-intensity maps of the representative molecular lines for IRAS 15398-3359. Each column is labeled with the respective molecular line. On the left, we show the large-scale structure of $^{12}$CO (**a**). On the right, we show C$^{18}$O, $^{13}$CO and SO, with a large-scale view in the top row (**b**, **d**, and **e**) and a zoomed-in view in the bottom row (**c**, **e**, and **g**), respectively. The yellow star represents the protostellar position. The beamsizes are shown as the gray circles in the bottom left of each map, and scalebars are shown in the bottom right of the $^{12}$CO map, as well as the right-most SO maps for different physical scales. The solid and dashed gray lines indicate the deconvolved position angle of the continuum major and minor axes, respectively. The blue and red arrows represent the overall blue and red-shifted outflow directions found by Vazzano et al. (2021), respectively. The black contours show the integrated-intensity in steps 3, 5, 7, 9, 12, 15, 20, 50 and 100$\sigma_{\rm rms}$, where $\sigma_{\rm rms}$ is the rms noise of the respective integrated-intensity map.

### 3.2.2. *C$^{18}$O (2–1)*

The large-scale view of C$^{18}$O is shown in Figures 3b & 4b, while a zoomed-in view of the inner region is shown in Figures 3c & 4c. The C$^{18}$O emission is more centralized, with an asymmetric ring-like structure surrounding the central source (Figures 3b & 3c). There is a clear velocity gradient in the velocity map (Figure 4c) going from northeast (red) to south-west (blue), which is similar to the continuum minor axis and almost perpendicular to the $^{13}$CO and SO velocity gradients. The C$^{18}$O velocity gradient then twists towards the counter-clockwise direction and becomes more aligned to the continuum major axis at larger scales around the disk. The C$^{18}$O channel map shows the emission faintly traces the bipolar outflows, as well as some faint extended emission at low velocities perpendicular to the outflows (Figure B.2). In the SB-only integrated intensity map (Figures C.1b & C.1c), the asymmetric structure is not as clear as in the SB+LB maps. In the intensity-weighted velocity maps, the faint extended emission is more clearly seen (Figure C.2b). Additionally, the overall velocity gradient around the central source appears more consistent with the continuum major axis (Figure C.2c).

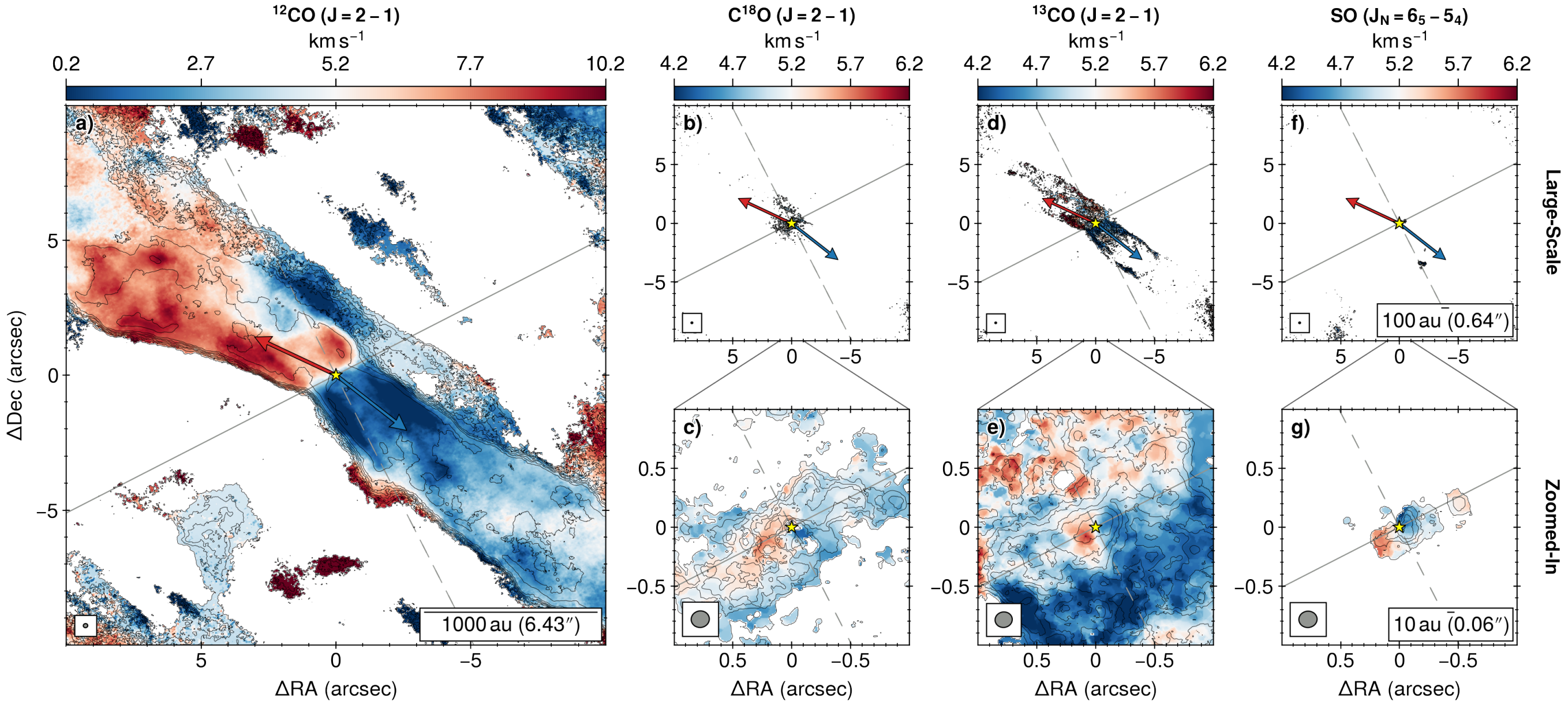

**Figure 4.** Same as Figure 3, but for the intensity-weighted velocity maps of the representative molecular lines.

### 3.2.3. *$^{13}$CO (2–1)*

The large-scale view of $^{13}$CO is shown in Figures 3d & 4d, while the inner region is shown in Figures 3e & 4e. The $^{13}$CO emission traces parts of the outflow cavity and inner structure around the protostar (Figures 3d & 3e). There is a clear velocity gradient around the source position that is consistent with the continuum major axis, indicating rotation around the center. The $^{13}$CO channel map shows a similar structure as C$^{18}$O, with the outflows and perpendicular extended emission both present (Figure B.3). In the $^{13}$CO channel map (Figure B.3), the south-east structure is clearly seen between 4.19 km s$^{-1}$ and 4.85 km s$^{-1}$, while the north-west is present from 4.52 km s$^{-1}$ and 4.85 km s$^{-1}$. Similar to $^{12}$CO, the velocity increases as the emission moves further away from the source. This is more clear in the south-east structure. In the SB-only maps (Figures C.1d & C.1e), the extended structures perpendicular to the outflow are clear and blue-shifted from the centroid velocity. The south-east structure extends out ∼800 au and reaches the blue-shifted patch of $^{12}$CO previously mentioned. The structure to the north-west is shorter and reaches out to a few hundred au. The extended structures are more clear in $^{13}$CO than in C$^{18}$O. In the SB+LB maps, these extended structures are resolved out.

### 3.2.4. *SO ($6_5$–$5_4$)*

The large-scale view of SO is shown in Figures 3f & 4f, while the inner region is shown in Figures 3g & 4g. The SO emission is compact and mostly located around the center near the protostellar position (Figures 3f & 3g). SO emission can either be enhanced in and around the protostellar disk due to accretion shocks from the infalling material (e.g., Sakai et al. 2014; Ohashi et al. 2014; Sakai et al. 2017; van Gelder et al. 2021) or MHD disk winds (Tabone et al. 2017). There is a clear velocity gradient that is consistent with both $^{13}$CO and the continuum major axis (Figure 4g). A higher excited transition of SO ($J_N = 7_6 \rightarrow 6_5$) was previously shown to trace rotation associated with a protostellar disk (Okoda et al. 2018). A small, offset patch of red-shifted emission is seen in the disk plane which could be associated with an accretion shock or the infalling envelope. There is a blue-shifted patch of SO emission that is coincident with the previously mentioned $^{12}$CO patch in the south-east at ∼ 1000 au. Also, there is a small patch of SO at the bottom edge of the blue-shifted outflow, which likely traces a shocked region in the outflow. The SB-only large-scale maps show the extended feature and the outflow shock more clearly (Figures C.1f & C.2f). While the zoomed-in map also shows a velocity gradient similar to that in the SB+LB map (Figures C.1g & C.2g).

## 4. ANALYSIS

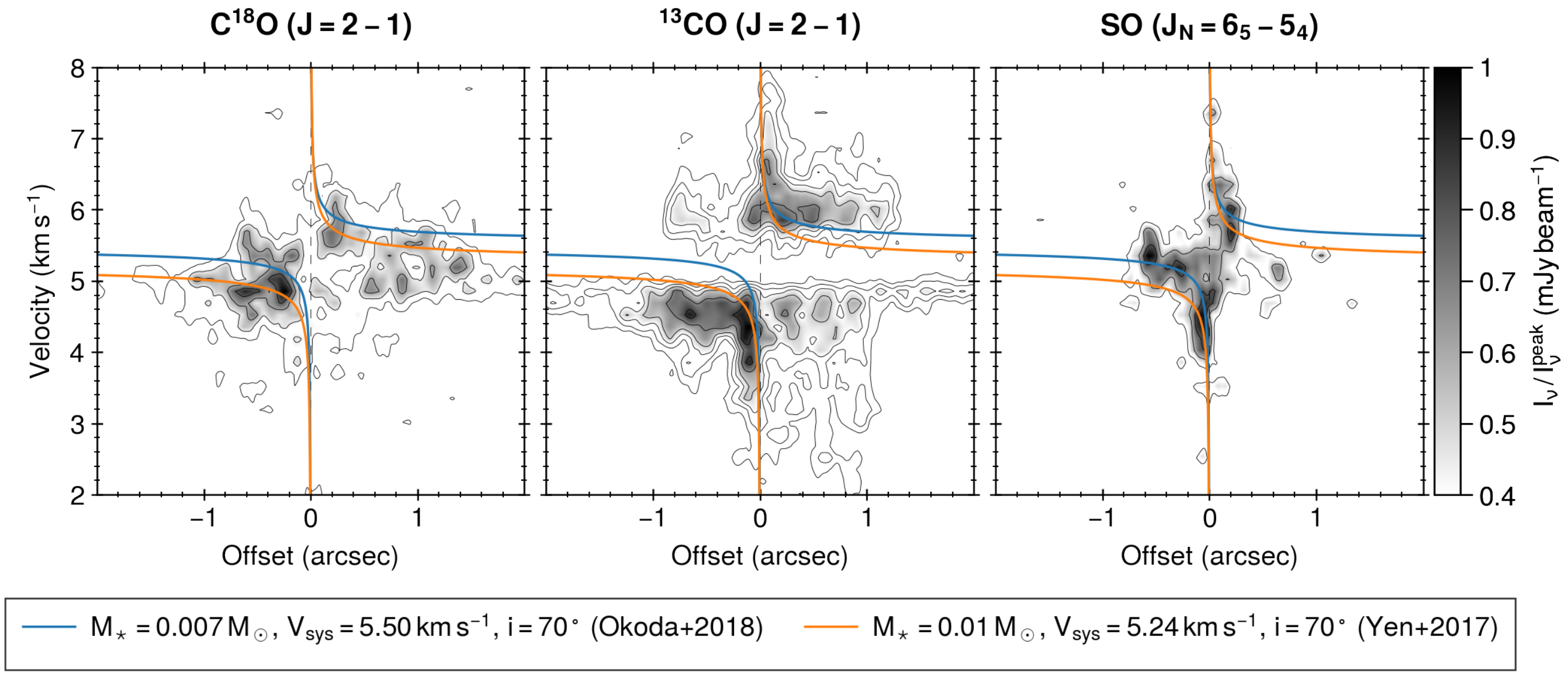

**Figure 5.** Position-Velocity diagrams of $C^{18}O$ (**left**), $^{13}CO$ (**center**), and SO (**right**). All cuts were made with a width of 1 beam. The offset axis goes from west (negative) to east (positive). The colorbar scale is normalized by the peak intensity ($I_\nu^{\rm peak}$). The emission is also shown with black contours in steps 3, 5, 7, 9, 12, 15, 20, 50 and $100\sigma_{\rm rms}$, where $\sigma_{\rm rms}$ is the rms noise of the respective PV map. The blue and orange lines represent the Keplerian curves for the protostellar masses, system velocities and disk inclination angles reported by Okoda et al. (2018) and Yen et al. (2017a), respectively.

### 4.1. *Position-Velocity Diagrams*

Our high angular resolution data allows us to investigate the Keplerian disk in this source at smaller scales than in previous studies. We make some initial position-velocity (PV) diagrams to get an overview of the velocity structure and look for lines that show a Keplerian rotation signature. We investigate $C^{18}O$, $^{13}CO$ and SO by making PV cuts with widths of 1 beam across the continuum disk major axis (PA=117.1°) as measured from the dust continuum fitting. The resulting PV diagrams are shown in Figure 5.

The $C^{18}O$ emission takes a diamond shape, with emission in all four quadrants of the PV diagram (Figure 5, left panel). This is indicative of infall dominated motions (e.g., Ohashi et al. 1997; Harsono et al. 2015). The PV diagram of $^{13}CO$ is much more complicated (Figure 5, middle panel). The red and blue-shifted velocity signature seen in the center of the integrated-intensity map (Figure 4e) likely corresponds to some of the emission seen in the upper right and lower left quadrants, which could be tracing infalling and rotational components of the system. Though, there are several high velocity structures in all four quadrants, which are likely due to contamination by the outflow kinematics. The lack of emission at velocities around the line center is likely due to filtering out by the high-angular resolution. The PV structure in the SO map seems the most promising to trace a possible disk rotation signature around the protostar (Figure 5, right panel). The overall emission mainly shows up in the top right and bottom left quadrants of the PV diagram. This is expected for rotational motions, where faster rotation would shift the emission peaks to smaller offsets (e.g., Harsono et al. 2015). This PV structure is also consistent with a higher transition of SO (at $\sim 261$ GHz) that was previously analyzed by Okoda et al. (2018).

We briefly compare the PV structures in Figure 5 to Keplerian rotation curves using previous protostellar mass estimations (Yen et al. 2017a; Okoda et al. 2018). We assume the system velocity to be $5.24\,{\rm km\,s^{-1}}$ as previously estimated by Yen et al. (2017a) using $C^{18}O$. It is worth noting that Okoda et al. (2018) employ a higher system velocity of $5.5\,{\rm km\,s^{-1}}$ for their Keplerian analysis of SO. In our case, the overall PV structure of $C^{18}O$ and $^{13}CO$ is more consistent with the lower value. The Keplerian curves seem to match well with the higher velocity SO emission near the source. As for $C^{18}O$ and $^{13}CO$, it is not clear whether the emission corresponds to the rotation curves in the same high velocity region near the source. Thus, SO seems to be the most promising for studying the dynamics of the protostellar disk in IRAS15398.

### 4.2. *PV Analysis with SLAM*

We employ the rotation curve fitting tool from the Spectral Line Analysis/Modeling (SLAM) python

**Table 4.** SLAM Fitting Results for SO ($J_N = 6_5 \rightarrow 5_4$)

| Fitting Description | Fitting Method | $R_b$ (au) | $p$ | $dp$ | $V_{\rm sys}$ ($\rm km\,s^{-1}$) | $M_b$ ($M_\odot$) | $\bar{\chi}^2$ |
|---|---|---|---|---|---|---|---|
| **Initial** | Edge | $41.39^{+1.12}_{-1.13}$ | $1.09^{+0.09}_{-0.08}$ | – | $5.42^{+0.02}_{-0.02}$ | $0.0299 \pm 0.0008$ | 3.8 |
| - $V_{\rm range} = 0.0 - 2.0\,\rm km\,s^{-1}$, $p$ = free, $V_{\rm sys}$ = free | Ridge | $57.84^{+2.64}_{-2.60}$ | $0.81^{+0.03}_{-0.03}$ | – | $5.47^{+0.01}_{-0.01}$ | $0.0066 \pm 0.0003$ | 6.3 |
| **Modified 1** | Edge | $23.05^{+1.81}_{-1.77}$ | $0.85^{+0.28}_{-0.18}$ | – | $5.37^{+0.08}_{-0.05}$ | $0.053 \pm 0.004$ | 1.6 |
| - $V_{\rm range} = 0.6 - 2.0\,\rm km\,s^{-1}$, $p$ = free, $V_{\rm sys}$ = free | Ridge | $10.38^{+0.93}_{-0.90}$ | $0.64^{+0.21}_{-0.11}$ | – | $5.44^{+0.08}_{-0.05}$ | $0.024 \pm 0.002$ | 1.4 |
| **Modified 2** | Edge | $19.40^{+1.12}_{-1.10}$ | 0.5 | – | $5.30^{+0.02}_{-0.02}$ | $0.045 \pm 0.003$ | 1.8 |
| - $V_{\rm range} = 0.6 - 2.0\,\rm km\,s^{-1}$, $p$ = fixed, $V_{\rm sys}$ = free | Ridge | $9.68^{+0.66}_{-0.64}$ | 0.5 | – | $5.40^{+0.03}_{-0.03}$ | $0.022 \pm 0.002$ | 1.2 |
| **Modified 3** | Edge | $24.12^{+1.37}_{-1.50}$ | 1.0 | – | $5.40^{+0.04}_{-0.04}$ | $0.055 \pm 0.003$ | 1.6 |
| - $V_{\rm range} = 0.6 - 2.0\,\rm km\,s^{-1}$, $p$ = fixed, $V_{\rm sys}$ = free | Ridge | $11.11^{+0.95}_{-0.98}$ | 1.0 | – | $5.57^{+0.06}_{-0.05}$ | $0.026 \pm 0.002$ | 2.3 |
| **Modified 4** | Edge | $21.30^{+49.08}_{-7.31}$ | $0.71^{+0.15}_{-0.32}$ | $0.36^{+0.49}_{-0.26}$ | $5.40^{+0.02}_{-0.02}$ | $0.06 \pm 0.1$ | 2.4 |
| - $V_{\rm range} = 0.2 - 2.0\,\rm km\,s^{-1}$, $p$ = free, $dp$ = free, $V_{\rm sys}$ = free | Ridge | $19.23^{+41.40}_{-13.39}$ | $0.74^{+0.05}_{-0.022}$ | $0.26^{+1.29}_{-0.20}$ | $5.46^{+0.01}_{-0.01}$ | $0.014 \pm 0.04$ | 8.3 |
| **Modified 5** | Edge | $18.25^{+4.82}_{-3.28}$ | 0.5 | 0.5 | $5.39^{+0.02}_{-0.02}$ | $0.07 \pm 0.03$ | 14.4 |
| - $V_{\rm range} = 0.2 - 2.0\,\rm km\,s^{-1}$, $p$ = fixed, $dp$ = fixed, $V_{\rm sys}$ = free | Ridge | $21.87^{+1.44}_{-1.60}$ | 0.5 | 0.5 | $5.45^{+0.01}_{-0.01}$ | $0.016 \pm 0.002$ | 4.7 |

Note— $R_b$ is the break radius, $p$ is the power-law index (or inner power-law index for the double power-law function), $dp$ is the change in the double power law index between the inner and outer functions, $V_{\rm sys}$ is the systemic velocity, $M_b$ is the dynamical mass at the break radius and $\bar{\chi}^2$ is the reduced chi-squared of the fit. For the single power-law fittings, the break radius is arbitrarily chosen from within the derived points. For the double power-law fitting, $p = p_{\rm in}$ and $dp = p_{\rm out} - p_{\rm in}$ and the break radius is where the power-law transitions from inner to outer. The uncertainties reported are the 16th and 84th percentiles of the posterior distributions in the MCMC fitting.

package[2], which uses Markov-Chain Monte Carlo (MCMC) sampling to fit a single or double power-law profile to the emission in the PV diagram (Aso & Sai 2023). Here, we present a brief overview of the fitting procedure and methodology and refer the reader to Ohashi et al. (2023) for a complete description (see also: Aso et al. 2015; Sai et al. 2020; Aso & Machida 2020). In the case of a single power-law, the velocity is described by

$$v = V_b \left(\frac{r}{R_b}\right)^{-p}, \tag{4}$$

where $R_b$ is an arbitrarily chosen "break" radius at a fixed velocity $V_b$, and $p$ is the power-law index. The double power-law case describes the velocity as

$$v = \begin{cases} V_b \left(\frac{r}{R_b}\right)^{-p_{\rm in}} & (r \leqslant R_b) \\ V_b \left(\frac{r}{R_b}\right)^{-p_{\rm out}} & (r > R_b) \end{cases} \tag{5}$$

where $p_{\rm in}$ and $p_{\rm out}$ are the inner and outer power-law indices, respectively. Instead of strictly defining $p_{\rm out}$, SLAM instead implements the variable $dp$, which is simply the change in power-law indices ($dp = p_{\rm out} - p_{\rm in}$). We will describe $p$ and $dp$, instead of $p_{\rm in}$ and $p_{\rm out}$, for the remainder of the paper to simplify the descriptions between single and double power-law indices. A power-law index of $p \approx 0.5$ represents a Keplerian rotating disk, while $p \approx 1$ represents infalling material with conserved angular momentum. The algorithm finds the edges and ridges of the emission in the PV diagram by cutting along the position and/or velocity axes. One can set an intensity threshold at which to find the edge and ridge points. Changing the threshold mainly affects the edge calculation, as the outermost radius with this value is the edge radius. For the ridge calculation, the mean is found using data above the threshold. The edge and ridge points are then separately fit with the power-law function to find the best parameters. The dynamical mass can then be estimated via the Kepler mass by

$$M = v^2 r / G / \sin^2 i, \tag{6}$$

where $G$ is the gravitational constant and $i$ is the source inclination angle. The mass can be interpreted as the dynamical protostellar mass ($M_\star$) when $p \approx 0.5$, and the true protostellar mass would then be somewhere between the break mass, $M_b$, estimated from the edge (upper mass limit) and ridge (lower mass limit) methods. If $p$ is not Keplerian but close to it, $M_b$ tells us a reasonable range of the protostellar mass. Additionally, the break radius only represents the disk radius when fitting the double power-law function. In the case of a single power-law, the break radius is arbitrarily chosen from within the derived points,

[2] The Spectral Line Analysis/Modeling (SLAM) tool is publicly available at https://github.com/jinshisai/SLAM. This tool was developed for the eDisk large program to derive dynamical protostellar masses.

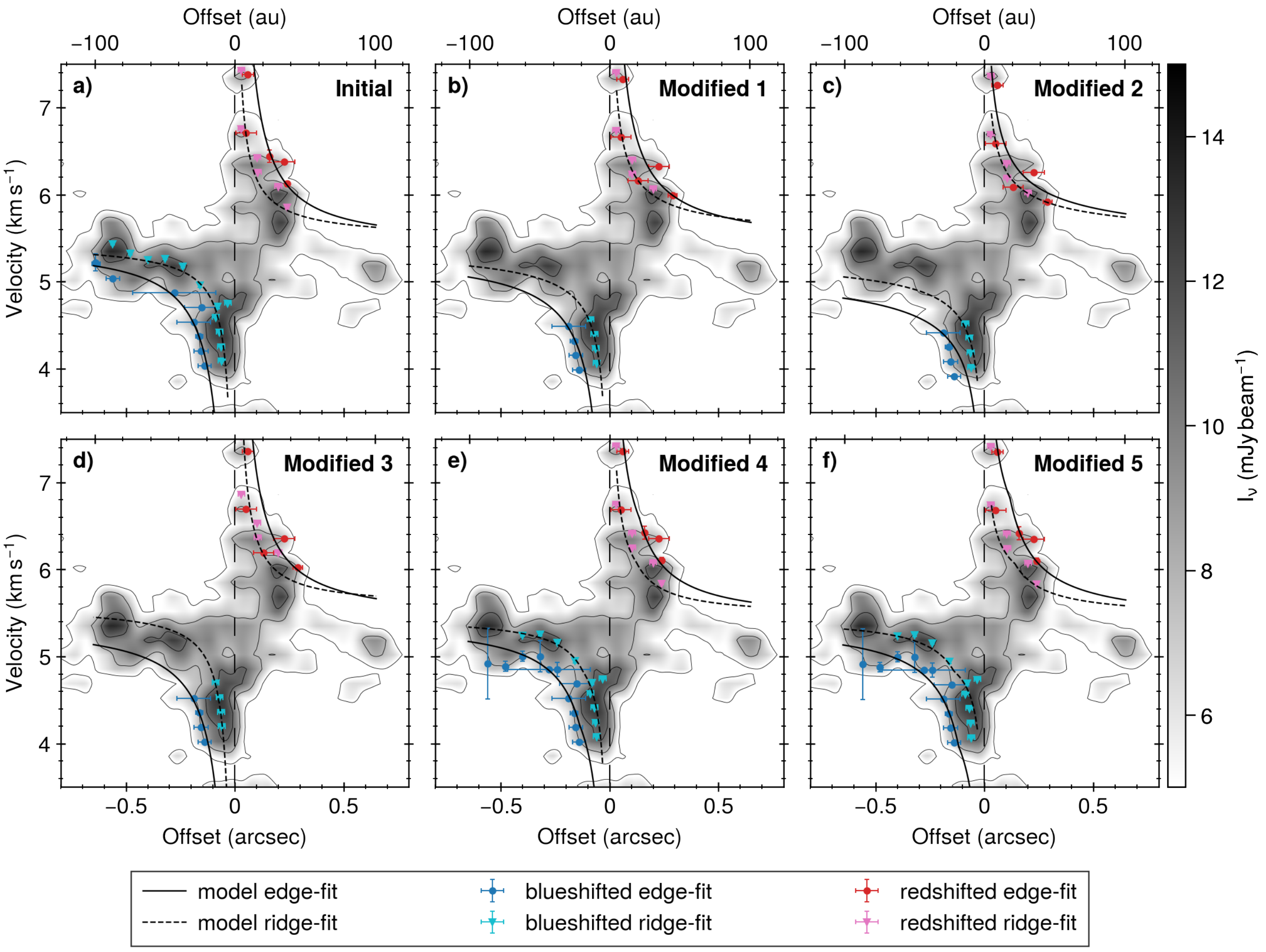

**Figure 6.** The Position-Velocity (PV) plots of the SLAM fitting results for SO. The red and blue circles represent the red and blue-shifted points fit with the edge method, while the magenta and cyan triangles represent the red and blue-shifted points fit with the ridge method. The black solid and dashed lines show the best fit edge and ridge models, respectively. The contours in the PV map are shown in steps 3, 5, 7, 9, 12, 15, 20, 50 and $100\sigma_{\rm rms}$, where $\sigma_{\rm rms}$ is the rms noise in the PV map.

usually where the cuts for both position and velocity meet.

In the following sections, we describe multiple different PV diagram fittings of the SO emission using single and double power-law profiles from SLAM. Our high-resolution data will provide the first PV diagram fitting of the IRAS15398 protostellar disk, as the previous observations by Yen et al. (2017a) did not resolve the disk. Fitting the PV diagram, especially for protostars with small disks, may not be very straight-forward which is shown by our multiple fitting attempts. We quantify the "goodness-of-fit" to assess which models best quantify the current data. Several parameters are kept the same among the different fits. We set the distance to IRAS15398 as 155.5 pc and the inclination angle to 50.7°, as estimated by the 2D Gaussian fitting in Section 3.1. We use a $5\sigma$ threshold, where the RMS of the PV diagram is measured to be $\sigma$ = 1.55 mJy beam$^{-1}$. A lower threshold of $3\sigma$ was also tested, however the reduced chi-squared was higher than when using a $5\sigma$ threshold. We explore values less than and $2\,M_\odot$ for the central protostellar mass, which is the expected limit given the evolutionary stage of the source as well as its luminosity. The SLAM fitting results are presented in Table 4. Additionally, we show the results in both Position-Velocity (PV) and Radius-Velocity (RV) space in Figures 6 and 7, respectively.

#### 4.2.1. *Single Power-Law*

**Initial:** We begin with an initial SLAM fitting using the single power-law function (Equation 4). The full position-velocity extent of the SO emission is utilized by setting the position range parameter from 0 to 100 au and the velocity range parameter from 0 to 2 km s$^{-1}$. Both position and velocity cuts are used to derive the edge and ridge points for fitting. The power-law index and systemic velocity are kept as free variables. The results of our initial fitting are shown in Figures 6a and 7a. The derived data points from the initial SLAM fit show no visual break in the slope, where one might expect a power-law transition from infalling material ($p \sim 1$) to disk ($p \sim 0.5$). This motivates our initial use of a single power-law function for fitting the data. At radii $< 10$ au, the slope of the points may start to flatten out around $0.3''$, but the lack of points at these smaller radii due to the angular resolution cannot confirm this. Curiously, there is one very low velocity point at $\sim$ 100 au. Including the systemic velocity as a free parameter gives a value of $\sim 5.45\,\mathrm{km\,s^{-1}}$. The best-fit power-law indices from the edge and ridge fits are both closer to $p \sim 1$, which indicates infalling material with conserved angular momentum. However, this result could be influenced by the low velocity points further away from the source, as well as the choice of the system velocity. The reduced chi-squared ($\bar{\chi}^2$) of the fits are 3.8 and 6.3 for the edge and ridge methods, respectively. It is important to see how these variables affect the overall fitting, and if it is possible to recover a Keplerian power-law profile.

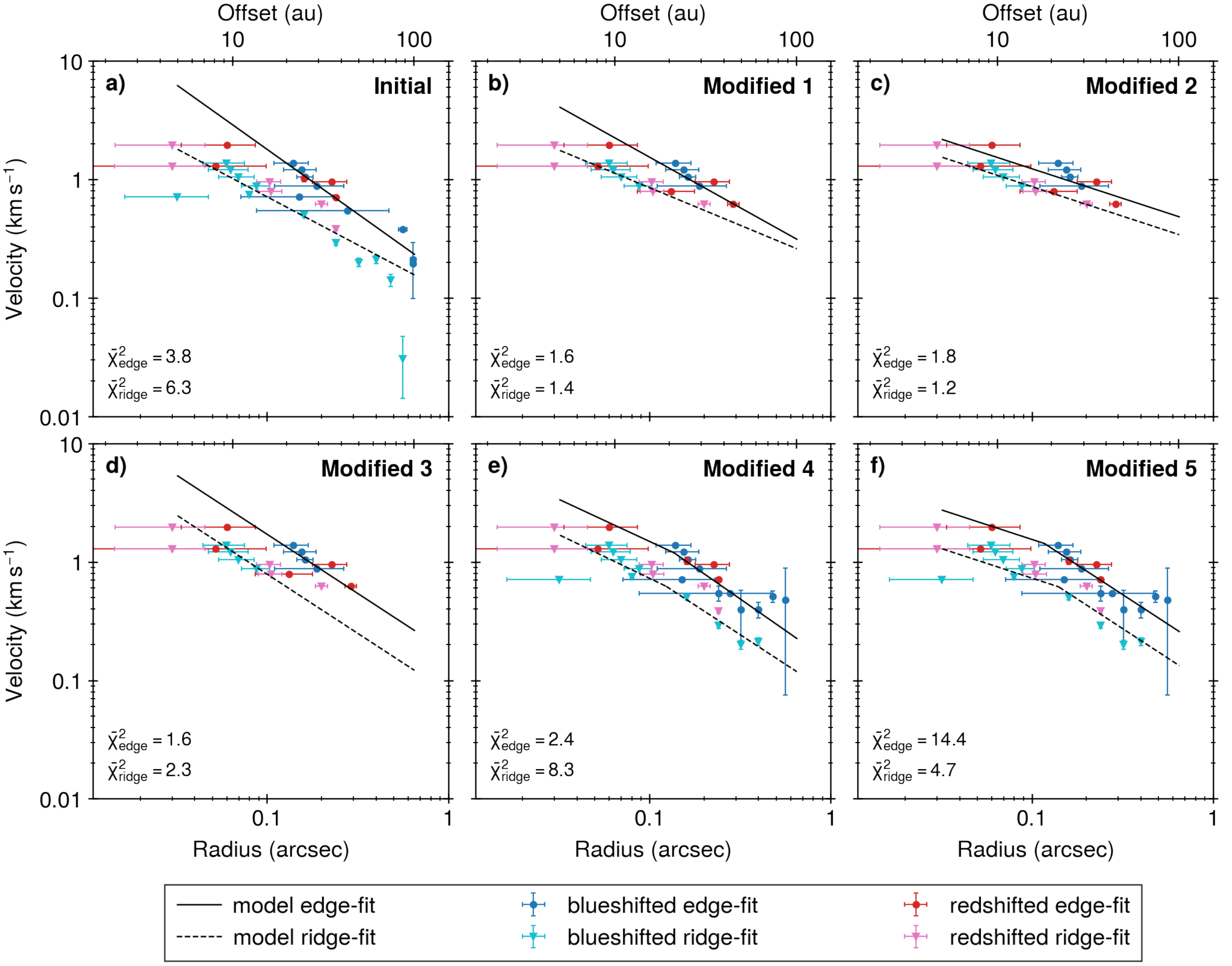

**Figure 7.** The Radius-Velocity (RV) plots of the SLAM fitting results for SO. The colored points and black lines have the same meaning as in Figure 6. The best-fit power-law equations are shown in the bottom left.

#### 4.2.2. *Modified Single Power-Law*

**Modified 1**: We first modify our initial SLAM fitting by increasing the lower limit of the velocity range to $0.6\,\mathrm{km\,s^{-1}}$, in order to only include the high velocity points in our fit. Additionally, we only fit along the velocity axis since peak positions in the PV emission are at similar offsets at these higher velocities. All of the other parameters remain the same. The results of this first modified fitting are shown in Figures 6b and 7b. The errors in the fitting results for the power-law index and system velocity are larger due to the limited number of data points at higher velocities included in the fit. The best-fit systemic velocity is $\sim 5.4\,\mathrm{km\,s^{-1}}$. This is slightly lower, but still consistent with the value derived from the initial fit. Overall, the power-law index between the edge and ridge methods is lower at around $p \sim 0.75$. The power-law index from the initial fit is within the upper error bound for both the edge and the ridge fits. On the other hand, the lower error bounds are now significantly closer to Keplerian rotation. Thus, it is difficult to make a concrete conclusion on whether the emission is related to the infalling envelope or a Keplerian disk. The $\bar{\chi}^2$ of the both the edge and ridge is lower than the initial fit at 1.6 and 1.4, respectively, indicating a better overall fit than the initial.

**Modified 2 & 3**: Since the power-law indices in our Modified 1 fit are somewhere between infall and Keplerian rotation, we fix the power-law indices to both scenarios ($p = 1.0$ and $p = 0.5$) to see how the $\bar{\chi}^2$ is affected in order to narrow down which scenario is more likely. We first run SLAM with a fixed $p = 0.5$ for Keplerian rotation (Modified 2). The $\bar{\chi}^2$ for the ridge method is 1.2, which is slightly lower than, but fairly consistent with the value in the Modified 1 ridge fit. This is due to the power-law index from the Modified 1 ridge estimate being closer to 0.5. In the edge method, the $\bar{\chi}^2$ is 1.8 and also consistent but slightly higher than in the Modified 1 edge fit. In contrast, when running SLAM with a fixed $p = 1.0$ for infalling envelope (Modified 3), the ridge method $\bar{\chi}^2$ becomes a much higher value of 2.3. This shows the ridge is tracing Keplerian rotation rather than infall, giving a lower-limit on the protostellar mass of $0.022 \pm 0.002, M_\odot$. The $\bar{\chi}^2$ for the edge method is actually the same value of 1.6 as in the Modified 1 fit. Thus, we cannot distinguish whether the edge follows Keplerian rotation or infall to constrain the protostellar mass upper limit. This could be due to insufficient angular resolution smearing the PV emission at the edge. The single power-law fittings do not directly give a disk radius, since the break radius from these fits is arbitrarily chosen to be within the derived points. However, we can infer a lower-limit of the protostellar disk radius using the outer most point of our ridge fit, which is $31.2 \pm 2.4$ au.

#### 4.2.3. *Double Power-Law*

**Modified 4 & 5**: The expected transition from infalling material to a Keplerian rotating disk provides a physical motivation to fit the observed PV diagrams with a double power-law function (Equation 5). Although a break in the data points is not clear, it is worth investigating if we can recover a potential protostellar disk radius and compare the $\bar{\chi}^2$ values to the single-power law fits. The first double power-law fit (Modified 4) leaves $p$ (or $p_{\rm in}$) and $dp$ as free parameters. Overall the inner and change in power-law indices are $p \sim 0.73$ and $dp \sim 0.31$ ($p_{\rm out} \sim 1.04$), which are very reasonable values and indicate the outer power-law index is more consistent with infall, while the inner power-law index is slightly higher than Keplerian. However, the $\bar{\chi}^2$ values of 2.4 and 8.3 for the edge and ridge, respectively, are much larger than the single power-law fits. This is also apparent in the Modified 5 fit, where $p$ and $dp$ are both fixed to 0.5 ($p_{\rm out} = 1.0$) and the $\bar{\chi}^2$ are also much higher at 14.4 and 4.7 for the edge and ridge. Comparing the two power-law fits, we do see a somewhat similar trend of the $\bar{\chi}^2$ for the edge and ridge as seen in the previous single power-law fits. The ridge method produces a lower $\bar{\chi}^2$ when $p$ and $dp$ are both fixed and the inner power-law is Keplerian, whereas the edge $\bar{\chi}^2$ is much higher for the fixed values and less when left free and the power-law is steeper than Keplerian. This really demonstrates the difficulty in fitting PV diagrams of the smallest protostellar disks, and caution should be taken when interpreting results.

## 5. DISCUSSION

### 5.1. *A Very Low-Mass Protostar and Disk*

The single power-law ridge fittings suggests the presence of a Keplerian disk. This constrains the lower limit of the dynamical protostellar mass of IRAS15398 to be $0.022 \pm 0.002\,M_\odot$. Unfortunately, the edge method in the single power-law fits does not clearly indicate existence of a Keplerian disk, since we cannot distinguish whether the power-law index is closer to 0.5 (Keplerian) and 1.0 (infall). Insufficient angular resolution could cause smearing near the edges of the PV diagram emission, making it difficult to recover the power-law index. Because of the unclear results of the PV edge in the single power-law fittings, it is not easy for us to estimate an upper limit of the stellar mass. Based on the outline of the edge contour of the PV diagram, we can conservatively estimate an upper limit protostellar mass to be $\sim 0.1\,M_\odot$. The double power-law fits are not used to determine the protostellar mass and disk radius due to the lack of clear break in the derived points used for fitting and the higher overall $\bar{\chi}^2$ values. The very low-mass indicates that IRAS15398 is likely in its infancy, which is also consistent with approximate ages (∼1000 years) derived from the dynamical timescale of the protostellar outflows (Bjerkeli et al. 2016a; Vazzano et al. 2021) and the lack of substructures in the dust disk. Earlier studies by Oya et al. (2014) and Yen et al. (2017a) used infalling envelope models to constrain the protostellar mass to be less than $0.09\,M_\odot$ and $0.01^{+0.02}_{-0.01}\,M_\odot$, respectively. It should be noted that the upper limit by Yen et al. (2017a) is under the assumption that the infall is free-fall, and if the infall is actually slower than this estimated mass would be a lower limit (see also Ohashi et al. 2014; Aso et al. 2015). These studies did not clearly detect a Keplerian rotation signature, but their derived protostellar masses are consistent with this study. Okoda et al. (2018) infer Keplerian rotation around a protostellar mass of $0.007^{+0.004}_{-0.003}\,M_\odot$ using a by-eye comparison to the molecular emission from their SO ($J_N = 7_6 \rightarrow 6_5$) data. The beamsize of their SO data is ∼1.5 times larger than ours at $0.22'' \times 0.16''$with an RMS noise ∼1.8 times higher at 4 mJy beam$^{-1}$ at a similar velocity resolution. Since the true protostellar mass likely lies between the edge and ridge fittings of the PV diagram, their comparison to the ridge provides only a lower limit on the protostellar mass. Additionally, Okoda et al. (2018) employ a disk inclination angle of 70°in their comparison, which is ∼ 20° higher than our estimate of 50.7° from 2D Gaussian fitting of the dust continuum. Adjusting the previous mass value for our new inclination angle increases the mass to $\sim 0.01\,M_\odot$, which is still consistent with our result as a lower limit. Overall, the method used for deriving the protostellar mass is inherently different. Okoda et al. (2018) perform a by-eye comparison using the ridge of the PV emission, which seemingly does match well to their PV structure. If a higher velocity of 5.5 km s$^{-1}$ is assumed, then the more extended, lower velocity emission on the blue-shifted side of the PV diagram could be interpreted as being in Keplerian rotation and thus change the resulting protostellar mass estimation when matching by-eye. When visually comparing our PV fitting results in Figure 6, it is somewhat difficult to tell the difference when the power-law is Keplerian vs. infall. Protostellar mass estimations using the methodology in this paper were tested with numerical simulations of protostellar disk formation at several epochs (Aso & Machida 2020). It was found that the true protostellar mass is likely between mass estimates from both the edge and ridge PV estimations. This means that using only the ridge method could potentially underestimate the true protostellar mass. However, Aso & Machida (2020) also show that the estimated mass from fitting their PV diagram was slightly larger than the actual protostellar mass when $M_\star < 0.4\,M_\odot$, due to the disk mass contributing to the overall rotational velocity in the very early stages. Since the protostellar mass of IRAS15398 is within this limit, it is possible to be slightly overestimated. Compared to other Class 0 sources, the protostellar mass of IRAS15398 is among the smallest (e.g., Tobin et al. 2012; Murillo et al. 2013; Ohashi et al. 2014; Yen et al. 2015b, 2017a; Aso et al. 2017; Maret et al. 2020). An analogous Class 0 protostar with a similarly small protostellar mass is B335 (Yen et al. 2015b). A protostellar mass of $\sim 0.05\,M_\odot$ was suggested from kinematic modeling, even though no Keplerian rotation was detected. These results hinted at the possibility of magnetic braking playing a crucial role in the evolution of B335. This aspect will be discussed more in Section 5.2, where we will compare the magnetic and Keplerian disk properties between IRAS15398 and B335.

The protostellar (gas+dust) disk mass is estimated to be $0.6-1.7\times10^{-3}\,M_\odot$ (or $0.6-1.8\,M_{\rm jup}$, $191-572\,M_\oplus$) using the integrated flux density from our 2D Gaussian fitting results under an optically thin assumption. If the emission at 1.3 mm is optically thick, the derived disk mass could only be a lower limit. Multi-band observations would be useful to constrain the optical depth at these wavelengths. Our assumed gas-to-dust ratio means that the dust-only disk mass is 100 times smaller at roughly $2-6\,M_\oplus$. The recent VANDAM surveys in Perseus and Orion provide estimates on the dust masses (and also dust disk radii; see Section 5.2) in a large sample of Class 0 sources (Segura-Cox et al. 2018; Tobin et al. 2020; Sheehan et al. 2022). From a sample of 14 Class 0 disk candidates, Segura-Cox et al. (2018) estimate the range of disk masses to be $0.01-0.46\,M_\odot$, which is a few orders of magnitude higher than the disk mass in IRAS15398. A more comprehensive study by Tobin et al. (2020) was performed that compared the dust disk masses of Class 0/I protostars in Orion to values of Class 0/I protostars in a few other clouds and also a sample of Class II sources. For Class 0 stars in Orion, the mean and median dust disk mass estimates are $25.9^{+7.7}_{-4.0}\,M_\oplus$ and $25.7^{+102.9}_{-6.7}\,M_\oplus$. The Class 0 protostars in Perseus exhibit overall higher mean and median dust disk masses of $376.5^{+220.3}_{-89.5}\,M_\oplus$ and $549.8^{+1149.0}_{-108.4}\,M_\oplus$. The dust mass is found to systematically decrease over time when compared to the dust disk mass values in Class II sources. For example, in Lupus, the mean and median protoplanetary dust disk masses are $5.08^{+1.78}_{-1.41}\,M_\oplus$ and $3.5^{+15.2}_{-0.8}\,M_\oplus$ (Ansdell et al. 2016). Radiative transfer modeling by Sheehan et al. (2022) has shown that for the Class 0 sample of protostars in Orion and showed that previous modeling by Tobin et al. (2020) may have overestimated the dust disk mass, with larger discrepancies in the lower mass dust disks. Thus, more detailed radiative transfer modeling of the dust continuum in IRAS15398 would be helpful to constrain the dust disk mass available for future planet formation.

#### 5.1.1. *Disk-to-Stellar Mass Ratio*

The disk-to-star mass ratio is used as an indicator of the gravitational stability of a protostar-disk system, where systems with $M_{\rm disk}/M_\star \gtrsim 0.3$ are more unstable (e.g., Lodato & Rice 2004, 2005). Previous surveys indicate typical values of $M_{\rm disk}/M_\star$ to be between 0.001 to 0.1 in more evolved sources (e.g., Manara et al. 2022, and references therein). Using the mean values of the protostellar disk mass ($0.61 - 1.69 \times 10^{-3}\,M_\odot$) and the lower-limit protostellar mass ($0.022\,M_\odot$), the upper-limit of the disk-to-star mass ratio is $\sim 0.06$. As mentioned previously, this value could be underestimated if the 1.3 mm dust emission is optically thick. However, the value of $M_{\rm disk}/M_\star$ is well within the limit of a gravitationally stable disk.

#### 5.1.2. *Mass-Accretion Rate*

The mass-accretion rate onto the protostar can be estimated by

$$\dot{M}_{\rm acc} = \frac{L_{\rm acc} R_\star}{G M_\star}, \tag{7}$$

where $L_{\rm acc}$ is the accretion luminosity, $R_\star$ is the radius of the protostar, and $G$ is the gravitational constant. We take $L_{\rm acc} \approx L_{\rm bol} = 1.4\,L_\odot$ (assuming $L_{\rm acc} >> L_\star$) to obtain an upper limit on the mass-accretion rate, since $L_{\rm bol} = L_{\rm acc} + L_\star$. The radius of the protostar is assumed to be $R_\star = 3\,R_\odot$ (Stahler et al. 1980). Using the lower-limit derived protostellar mass of $M_\star = 0.022\,M_\odot$ gives an approximate mass-accretion rate of $6.1 \times 10^{-6}\,M_\odot\,{\rm yr}^{-1}$, while our conservative upper limit gives $1.3 \times 10^{-6}\,M_\odot\,{\rm yr}^{-1}$. This mass accretion rate is consistent with the most recent estimate of $0.2 - 7.0 \times 10^{-6}\,M_\odot\,{\rm yr}^{-1}$, which were found using a combination of the dynamical timescale of the outflow and protostellar mass (Oya et al. 2014; Bjerkeli et al. 2016a; Okoda et al. 2018). Assuming the mass-accretion rate is constant, we can estimate the age of the protostar by $M_\star/\dot{M}_{\rm acc}$, which yields an age of approximately $0.4 - 7.5 \times 10^4$ yr, which would be a lower limit. However, IRAS15398 was shown to have undergone an accretion burst in the past $100{-}1000$ yr (Jørgensen et al. 2013). This means the accretion rate is not constant and this method of approximating age is more uncertain. The dynamical ages of the outflows are younger at $< 1000$ yr for both the red and blue-shifted lobes (Bjerkeli et al. 2016a; Vazzano et al. 2021), which could also be a consequence of the accretion burst.

Early theoretical modeling predicts that a typical mass accretion rate of protostars lies somewhere between $10^{-6} - 10^{-5}\,M_\odot\,{\rm yr}^{-1}$ (Hartmann et al. 1997), while magnetohydrodynamic (MHD) simulations predict values between $10^{-7}{-}10^{-4}\,M_\odot\,{\rm yr}^{-1}$, with higher mass-accretion rates at earlier times in the simulations (e.g., Machida & Matsumoto 2012). This further points towards IRAS15398 being a younger protostar, since its mass-accretion rate is on the higher end of values predicted from simulations.

#### 5.1.3. *A Proto-Brown Dwarf Candidate?*

The upper mass limit of a brown dwarf is $\sim 0.075\,M_\odot$ (Kumar 1963a,b; Hayashi & Nakano 1963; Burrows et al. 1997). Since the protostellar mass of IRAS15398 is currently within this regime, it is interesting to discuss the possibility of whether IRAS15398 will become massive enough to sustain the nuclear fusion of hydrogen in its core. The envelope mass around IRAS15398 has been previously estimated to be between $0.5\,M_\odot$ and $1.2\,M_\odot$ (Kristensen et al. 2012; Jørgensen et al. 2013). The protostellar core-to-star efficiency is estimated to be $\sim 30\%$ from both observations of starless cores and numerical simulations (Enoch et al. 2008; Offner & Chaban 2017). This efficiency would give a final lower-limit stellar mass estimation of $0.15\,M_\odot$ for IRAS15398, well above the mass limit of a brown dwarf.

### 5.2. *An Extremely Small Protostellar Disk*

The lower-limit of the protostellar gas disk radius is $31.2 \pm 2.4$ au. A previous upper limit of the disk radius was found to be $20^{+50}_{-20}$ au by Yen et al. 2017a, which is within error of our estimate. The gas disk radius in IRAS15398 is comparable to, but on average smaller than that of other kinematically detected disks. The study by Yen et al. (2017a) looks at the properties of several Class 0/I protostars with Keplerian disks. The gas disk radii in their comparison ranges from 3 au to 150 au in the Class 0 stage. The protostellar disk is expected to continue to grow as more material and angular momenta from the natal, protostellar envelope collapses towards the central region. From a theoretical perspective, small Keplerian disks could indicate efficient magnetic braking of the infalling material (Allen et al. 2003; Hennebelle & Fromang 2008; Mellon & Li 2008b). The parsec-scale magnetic field in IRAS15398 has been previously studied using SOFIA polarization observations (Redaelli et al. 2019). They conclude that the core is strongly magnetized with a uniform-to-turbulent magnetic ratio of 0.27. Additionally, the overall magnetic field direction appears well-aligned to the outflow direction. Numerical MHD simulations have inversely found that increasing the misalignment between the magnetic field and rotation axis can help to overcome magnetic braking and promote disk formation (e.g., Hennebelle & Ciardi 2009; Li et al. 2013; Joos et al. 2012; Krumholz et al. 2013; Hirano et al. 2020). Thus, this alignment could indicate more efficient magnetic braking which would inhibit the growth of a large, protostellar disk.

The dust disk radius is taken to be the $2\sigma$ width of the deconvolved major axis ($\mathrm{FWHM_{maj}} = 2.355\sigma$) of our 2D Gaussian continuum fitting, which is 3.82 au. In the aforementioned VANDAM survey of Perseus, the dust disk radii ranged between 10.5 au and 42.2 au (Segura-Cox et al. 2018). In Orion, the mean and median dust disk radii were found to be $44.9^{+5.8}_{-3.4}$ au and $48.1^{+79.6}_{-24.5}$ au (Tobin et al. 2020). These values are consistently larger than the dust disk in IRAS15398, possibly hinting at other mechanisms, like environment effects, that are keeping its dust disk small.

#### 5.2.1. *Dust vs. Gas Disk Radius*

Small dust disks could be a signpost of the radial drift of dust grains as they grow in size and decouple from the gas disks. Discrepancies between the dust and gas disk radii have been observed in more evolved protoplanetary disks (e.g., Panić et al. 2009; Andrews et al. 2012; de Gregorio-Monsalvo et al. 2013; Ansdell et al. 2018; Trapman et al. 2020). Recent work by Ansdell et al. (2018) has revealed the radii of gas disks were $\sim$ 1.5 to 3.0 times that of the dust disks in their sample of 22 protoplanetary disks in Lupus. We report that the gas disk radius in IRAS15398 is $> 8$ times larger than the dust disk radius, which is significantly larger than later stage disks. In the early, deeply embedded protostellar disk stage, there are currently no studies that have investigated this before the eDisk program. However, if planet formation begins in these early stages, as suggested by substructure formation in a few Class I protostars (ALMA Partnership et al. 2015; Sheehan & Eisner 2018; Segura-Cox et al. 2020; Yamato et al. 2023), then this value could be essential for constraining the dust growth and radial drift in Class 0/I protostars. It should be noted, that observed differences in gas and dust disk radii could also be observed due to the optical depth effects (e.g., Facchini et al. 2017). Further modeling to constrain the optical depths is needed to disentangle these scenarios.

#### 5.2.2. *Specific Angular Momentum*

The specific angular momentum can be calculated via

$$j = \sqrt{GM_\star R_d}, \tag{8}$$

where the lower limit gas disk radius, $R_d \geq 31.2$ au, is used. At the lower-limit of the gas disk, the specific angular momentum is found to be $1.2 \times 10^{-4}\,\mathrm{km\,s^{-1}}$ pc. Using our conservative upper limit of $0.1\,M_\odot$ and the upper error limit gas disk radius $R_d \approx 70$ au from Yen et al. (2017a), the maximum specific angular momentum is $3.8 \times 10^{-4}\,\mathrm{km\,s^{-1}}$ pc. Previous values of the specific angular momentum found at larger scales of 140 au and 600 au were found to be $\sim 7 \times 10^{-5}\,\mathrm{km\,s^{-1}}$ pc and $\sim 1 \times 10^{-4}\,\mathrm{km\,s^{-1}}$ pc, respectively (Yen et al. 2017a).

While the uncertainty from the mass range is so large, the specific angular momentum seems to increases at smaller radii near the disk. One would expect the specific angular momentum to increase near the disk due to the dependence on radius. In the inside-out collapse scenario, the angular momentum profile is shown to flatten out in the inner $\sim$1000 au (Shu 1977; Yen et al. 2011). Assuming a constant angular momentum profile in the inner envelope down to the disk, an estimation of age can be inferred using the method described by Takahashi et al. (2016). Using Equation 26 in their paper, with specific angular momenta of $1.2 - 3.8 \times 10^{-5}\,\mathrm{km\,s^{-1}}$ pc, the age of IRAS15398 is expected to be $0.6 - 1.9 \times 10^4$ yr. This is consistent with the age calculated using the mass-accretion rate range in Section 5.1.2. These multiple age estimations, small protostellar mass and seemingly small disk demonstrate IRAS15398 is likely in its very early stages.

### 5.3. *A Dynamical Overview of the IRAS 15398-3359 Protostellar System*

IRAS15398 is a very young, low mass protostellar system hosting a small protostellar disk and two distinct bipolar outflows. We present a dynamical overview of the IRAS15398 system in Figure 8. We show the episodic outflow ejection lobes (Vazzano et al. 2021), the mean B-field direction (Redaelli et al. 2019), the linear structure interpreted either as a "relic" outflow (Okoda et al. 2021) or perturbed gas due a nearby possible outflow (Vazzano et al. 2021), as well as the rotation direction and PA of the disk from our study (Figure 8, left). Also, the previously derived outflow inclination angles (Oya et al. 2014; Yen et al. 2017a) are compared to the inclination angle found using the geometry of our 2D Gaussian dust continuum fitting (Figure 8, right).

Our results show that the PA of the continuum emission is not quite perpendicular to the overall outflow direction found by Vazzano et al. (2021). This could likely be related to the precessing protostellar outflow (Bjerkeli et al. 2016a; Vazzano et al. 2021). The precession is also seen in our $^{12}$CO data (Section 3.2.1), where the outflow emission bends in the plane of sky. We quantify this by calculating the difference in the continuum PA between the outflow PAs for the outer and inner episodic ejection lobes found by Vazzano et al. (2021). For the outer red and blue-shifted outflow lobes with PAs of 69° and 246°, the offset from the continuum major axis is ∼48° and ∼51°, respectively. For the inner red and blue-shifted outflow lobes with PAs of 37° and 214°, the offsets are closer to perpendicular at ∼80° and ∼83°, indicating the central dust emission aligns more perpendicularly with the more recent outflow ejections. One might expect in a perfectly symmetrical collapse that the outflows are perpendicular to the disk major axis. Thus, the offsets indicate the entire system is precessing. Since no binary is resolved, asymmetric infall seems a more likely scenario for causing the precession. We also note that the mean B-field direction is offset by ∼18° from the continuum minor axis. This could help to promote disk formation as found in previous MHD simulations where the angular momentum and B-field directional axes are misaligned (e.g., Krumholz et al. 2013; Li et al. 2013; Gray et al. 2018; Tsukamoto et al. 2018; Hirano et al. 2020; Wang et al. 2022). As for the inclination angles, although they seem to be different between the outflows and the disk, understanding the precession of the system in the plane-of-sky is less robust.

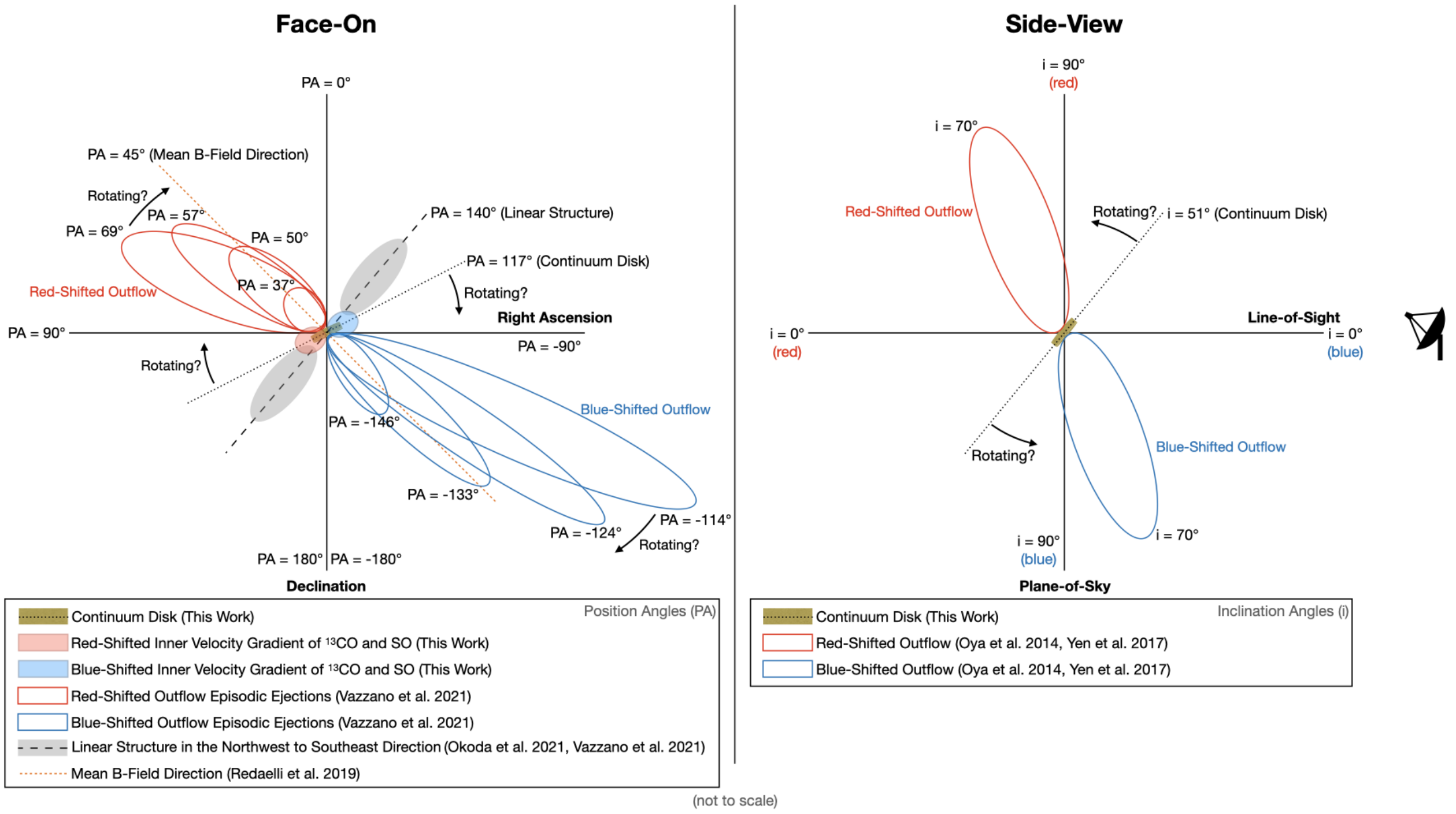

**Figure 8.** A dynamical schematic of the IRAS 15398-3359 protostellar system.

## 6. CONCLUSION

We have observed the Class 0 protostar, IRAS 15398-3359, at high-angular resolution with ALMA as part of the eDisk large program. This paper provides an initial analysis to derive various dynamical properties of the protostar-disk system. The main results are as follows:

1. The dust continuum observations show the extended dust structures as found in previous studies, but also reveal the small, compact dust disk at a high angular resolution ($\sim$ 0.04″). The compact dust emission indicates IRAS15398 is likely a single source, but still could be an extremely close binary with a separation of $\lesssim 6$ au. From the 2D Gaussian fitting, the $2\sigma$ radius of the dust disk is measured to be $\sim$ 3.8 au, with a position and inclination angle of 117.1° and 50.7°, respectively. The disk mass is found to be extremely low, with a value between $0.6-1.8\,M_{\rm jup}$. Observations at longer wavelengths may be needed to further constrain the disk mass if the dust is optically thick at 1.3 mm. No substructures within the dust disk were found. The molecular lines trace the gas kinematics of the bipolar outflows, inner envelope, and a rotating disk-like structure. The outflow is traced mainly by $^{12}$CO, and partially by $^{13}$CO in the SB+LB observations, while extended emission perpendicular to the outflows present in the SB-only maps. C$^{18}$O traces components of the inner envelope, while $^{13}$CO and SO show a rotational velocity gradient around the central protostar. The PV diagrams indicate the velocity structure of SO is likely tracing the rotating disk.

2. We employ SLAM to perform several rotation curve fittings using both single and double power-law profiles. We conclude from the $\bar{\chi}^2$ values of the single power-law fittings that the ridge traces a Keplerian rotating disk. This results in lower limit estimates for the protostellar mass and gas disk radius of $0.022 \pm 0.002\,M_\odot$ and $31.2 \pm 2.4$ au, respectively. The $\bar{\chi}^2$ for the single power-law edge fittings are less clear, but a conservative upper limit of the protostellar mass is $\sim 0.1\,M_\odot$. Overall, the reduced chi-squared of the single power-law fits are better than the double power-law fits.

3. We estimate the mass-accretion rate onto the protostar to be $1.3 - 6.1 \times 10^{-6}\,M_\odot\,\mathrm{yr}^{-1}$ and the specific angular momentum at the edge of the protostellar disk to be $1.2 - 3.8 \times 10^{-4}\,\mathrm{km\,s^{-1}\,pc}$. The low mass-accretion rate implies that there is still a significant amount of mass that can feed the star-disk system. From these two independent measurements, we estimate the age of the protostar is at least 3,600 years, all the way up to 75,000 years. At this young age with no clear substructures in the disk, planet formation would likely not yet have started. We estimate a disk-to-stellar mass ratio of <0.06, meaning the protostar-disk system is gravitationally stable. Additionally, the gas disk measured from our SLAM fitting of SO is >4.2 times larger than the dust disk measured from the dust continuum fitting. This could either be a signpost of dust radial drift or an optical depth effect. Further modeling of the dust continuum and molecular line optical depths could help to confirm which scenario is present in IRAS15398.

## ACKNOWLEDGMENTS

We thank the anonymous referee for their helpful comments and suggestions on this manuscript. This paper makes use of the following ALMA data: ADS/JAO.ALMA#2019.1.00261.L, ADS/JAO.ALMA#2019.A.00034.S. ALMA is a partnership of ESO (representing its member states), NSF (USA) and NINS (Japan), together with NRC (Canada), NSTC and ASIAA (Taiwan), and KASI (Republic of Korea), in cooperation with the Republic of Chile. The Joint ALMA Observatory is operated by ESO, AUI/NRAO and NAOJ. The National Radio Astronomy Observatory is a facility of the National Science Foundation operated under cooperative agreement by Associated Universities, Inc. This work used high-performance computing facilities operated by the Center for Informatics and Computation in Astronomy (CICA) at National Tsing Hua University. This equipment was funded by the Ministry of Education of Taiwan, the National Science and Technology Council of Taiwan, and National Tsing Hua University. SPL and TJT acknowledge grants from the National Science and Technology Council (NSTC) of Taiwan 106-2119-M-007-021-MY3 and 109-2112-M-007-010-MY3. N.O. acknowledges support from National Science and Technology Council (NSTC) in Taiwan through the grants NSTC 109-2112-M-001-051, 110-2112-M-001-031, 110-2124-M-001-007, and 111-2124-M-001-005. J.J.T. acknowledges support from NASA XRP 80NSSC22K1159. J.K.J. and R.S. acknowledge support from the Independent Research Fund Denmark (grant No. 0135-00123B). ZYL is supported in part by NASA NSSC20K0533 and NSF AST-2307199 and AST-1910106. LWL acknowledges support from NSF AST-2108794. H.-W.Y. acknowledges support from the National Science and Technology Council (NSTC) in Taiwan through the grant NSTC 110-2628-M-001-003-MY3 and from the Academia Sinica Career Development Award (AS-CDA-111-M03). S.T. is supported by JSPS KAKENHI Grant Numbers 21H00048 and 21H04495. This work was supported by NAOJ ALMA Scientific Research Grant Code 2022-20A. ZYDL acknowledges support from NASA 80NSSCK1095, the Jefferson Scholars Foundation, the NRAO ALMA Student Observing Support (SOS) SOSPA8-003, the Achievements Rewards for College Scientists (ARCS) Foundation Washington Chapter, the Virginia Space Grant Consortium (VSGC), and UVA research computing (RIVANNA). JEL is supported by the National Research Foundation of Korea (NRF) grant funded by the Korean government (MSIT) (grant number 2021R1A2C1011718). C.W.L. is supported by the Basic Science Research Program through the National Research Foundation of Korea (NRF) funded by the Ministry of Education, Science and Technology (NRF- 2019R1A2C1010851), and by the Korea Astronomy and Space Science Institute grant funded by the Korea government (MSIT; Project No. 2022-1-840-05). JPW acknowledges support from NSF AST-2107841. S.N. acknowledges support from the National Science Foundation through the Graduate Research Fellowship Program under Grant No. 2236415. Any opinions, findings, and conclusions or recommendations expressed in this material are those of the authors and do not necessarily reflect the views of the National Science Foundation. W.K.

was supported by the National Research Foundation of Korea (NRF) grant funded by the Korea government (MSIT) (NRF-2021R1F1A1061794). IdG acknowledges support from grant PID2020-114461GB-I00, funded by MCIN/AEI/10.13039/501100011033. Y.Y. is supported by the International Graduate Program for Excellence in Earth-Space Science (IGPEES), World-leading Innovative Graduate Study (WINGS) Program of the University of Tokyo. Y.A. acknowledges support by NAOJ ALMA Scientific Research Grant code 2019-13B, Grant-in-Aid for Scientific Research (S) 18H05222, and Grant-in-Aid for Transformative Research Areas (A) 20H05844 and 20H05847.

# APPENDIX

## A. GALLERY OF DUST CONTINUUM IMAGES

In Figure A.1, we present the short+long baseline dust continuum maps produced by the eDisk calibration and imaging script for IRAS15398. The properties of each map are shown in Table A.1. There is clear extended dust emission in the lower resolution maps, while increasing the angular resolution shows the extended emission becomes more resolved out until there is just a compact continuum structure in the highest resolution maps. The compact structure does appear to be disk-like, but is somewhat asymmetric. IRAS15398 does not show any signs of binarity up to our highest resolution of 5.8 au.

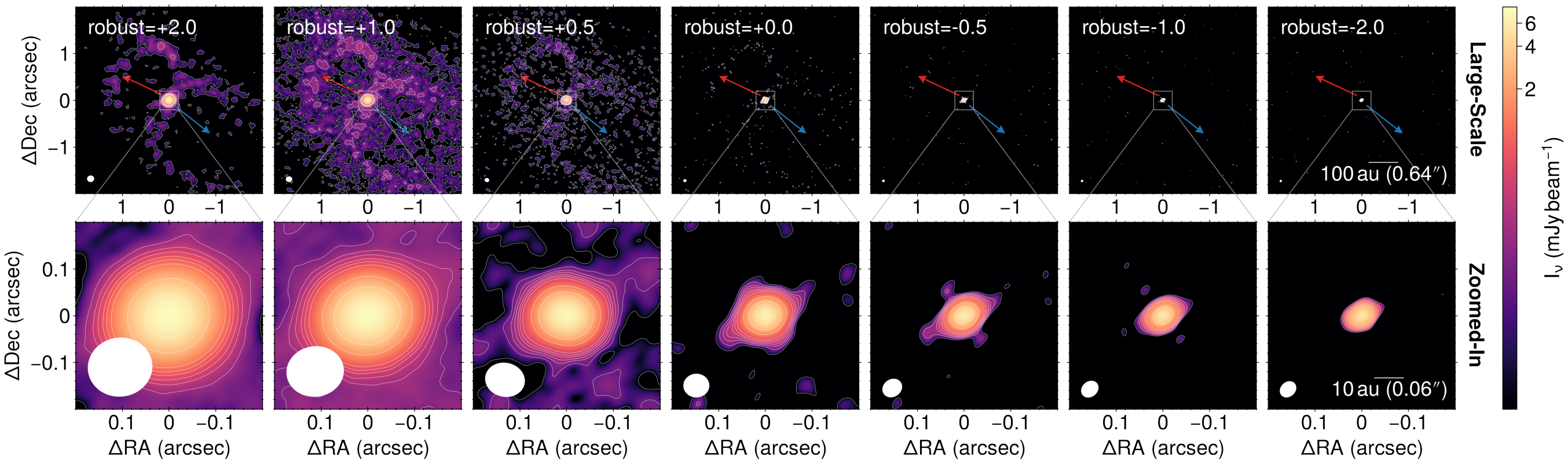

**Figure A.1.** Gallery of SB+LB continuum images from our eDisk observations of IRAS15398 with asinh (a=0.001) scaling. Each robust value made by the imaging script is shown from largest angular resolution (robust=+2.0) on the left, to smallest angular resolution (robust=-2.0) on the right. A large-scale view is shown in the top row, while a zoomed-in view is shown in the bottom row. The blue and red arrows represent the overall blue and red-shifted outflow directions found by Vazzano et al. (2021), respectively. The contours in both maps are shown in steps 3, 5, 7, 9, 12, 15, 20, 50 and 100$\sigma_{\rm rms}$, where $\sigma_{\rm rms}$ is the rms noise of each map.

**Table A.1.** Summary of SB+LB Dust Continuum Maps

| Robust | Beamsize | Peak Intensity (mJy beam$^{-1}$) | Peak Brightness Temperture (K) | RMS Noise (mJy beam$^{-1}$) | S/N Ratio |
|---|---|---|---|---|---|
| +2.0 | 0.″137 × 0.″125 (−81.8°) | 7.79 | 11.0 | 0.06 | 140 |
| +1.0 | 0.″122 × 0.″107 (−82.0°) | 7.63 | 14.2 | 0.03 | 237 |
| +0.5 | 0.″084 × 0.″071 (+78.0°) | 7.10 | 28.6 | 0.02 | 334 |
| +0.0 | 0.″054 × 0.″049 (−88.0°) | 6.47 | 58.9 | 0.03 | 248 |
| −0.5 | 0.″043 × 0.″036 (−55.8°) | 5.95 | 93.4 | 0.04 | 157 |
| −1.0 | 0.″038 × 0.″031 (−50.5°) | 5.68 | 114.2 | 0.06 | 98 |
| −2.0 | 0.″037 × 0.″030 (−47.4°) | 5.61 | 123.9 | 0.13 | 44 |

NOTE—Brightness temperatures were computed using the full blackbody equation at a frequency of 225 GHz. The peak intensity and brightness temperature of the continuum maps were found by using a 2″circular aperture around the center position. The rms noise of the continuum maps were found by using a 10″circular aperture in an emission-free area.

## B. CHANNEL MAPS OF MOLECULAR LINES

In Figures B.1-B.4, we present the short+long baseline molecular line channel maps for IRAS15398. Please see Section 3 for more description.

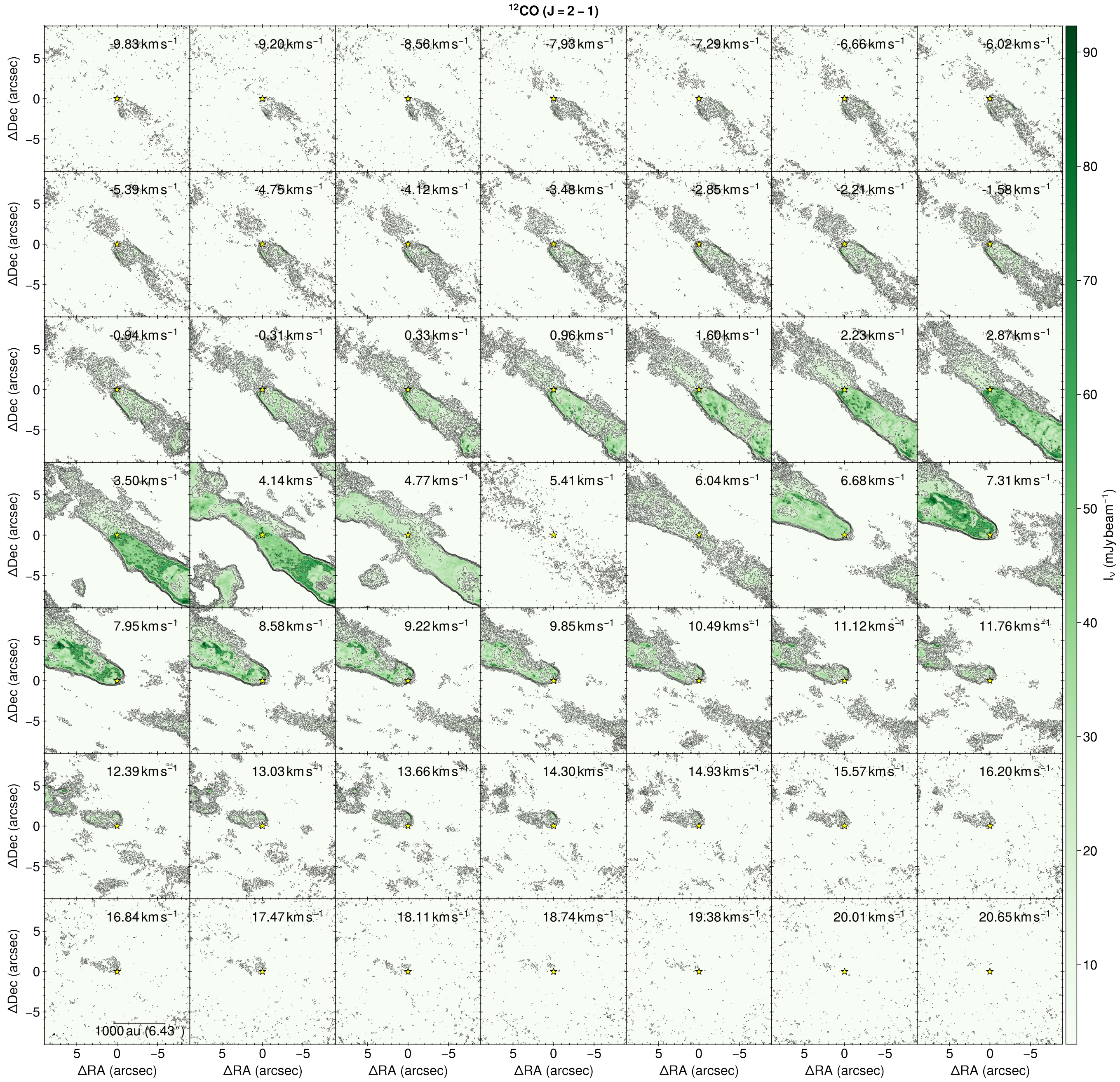

**Figure B.1.** Channel map of the SB+LB $^{12}$CO ($J = 2 \rightarrow 1$) molecular line data. The channel velocities are shown in the top right of the panels. The yellow star denotes the central position of IRAS15398 from the 2D Gaussian fitting. The scalebar is shown in the bottom left panel. Contours are shown in steps 3, 5, 7, 9, 12, 15, 20, 50 and 100$\sigma_{\rm rms}$, where $\sigma_{\rm rms}$ is the rms noise of the map. The channel map is centered at 5.41 km s$^{-1}$, which is the channel closest to the previously derived system velocity of 5.24 km s$^{-1}$ (Yen et al. 2017a).

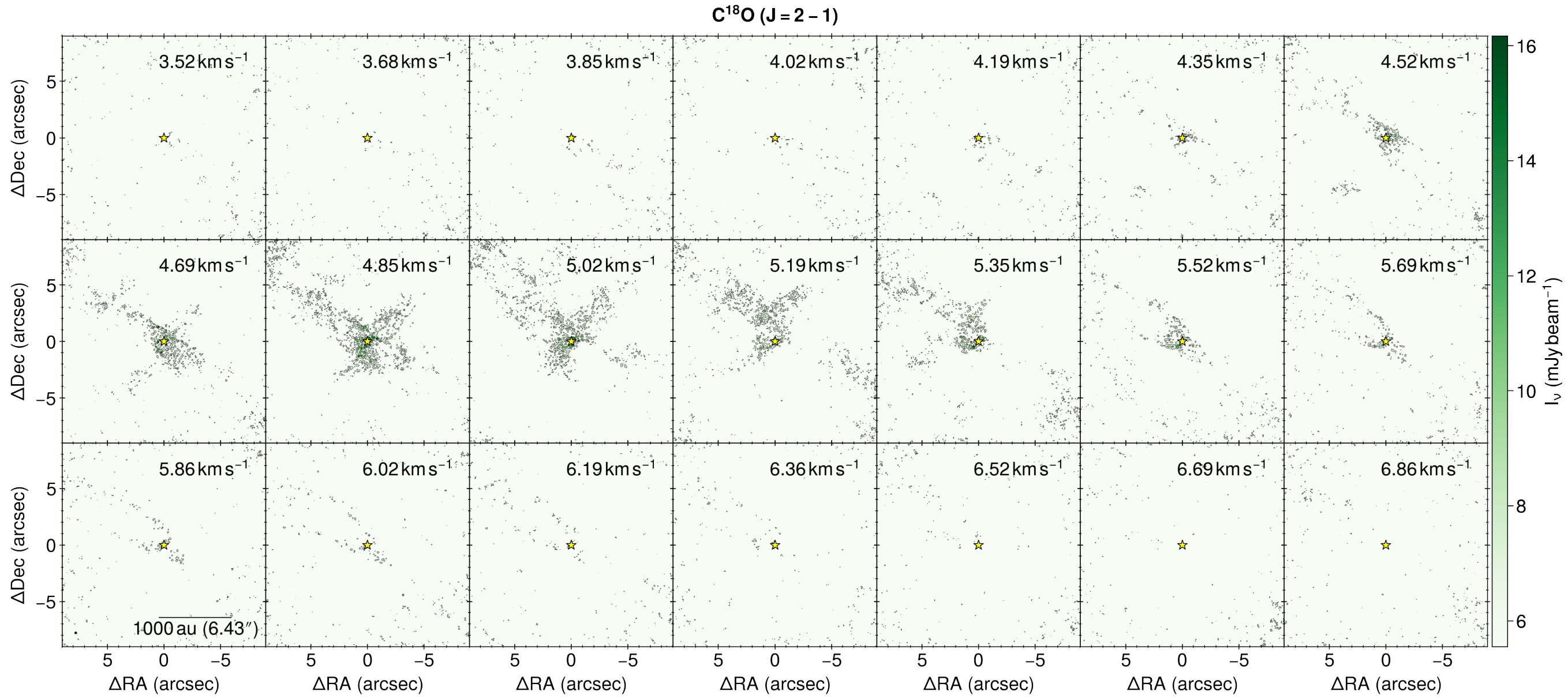

**Figure B.2.** Same as Figure B.1, but for $C^{18}O$ ($J = 2 \rightarrow 1$). The channel map is centered at $5.19\,\mathrm{km\,s^{-1}}$

.

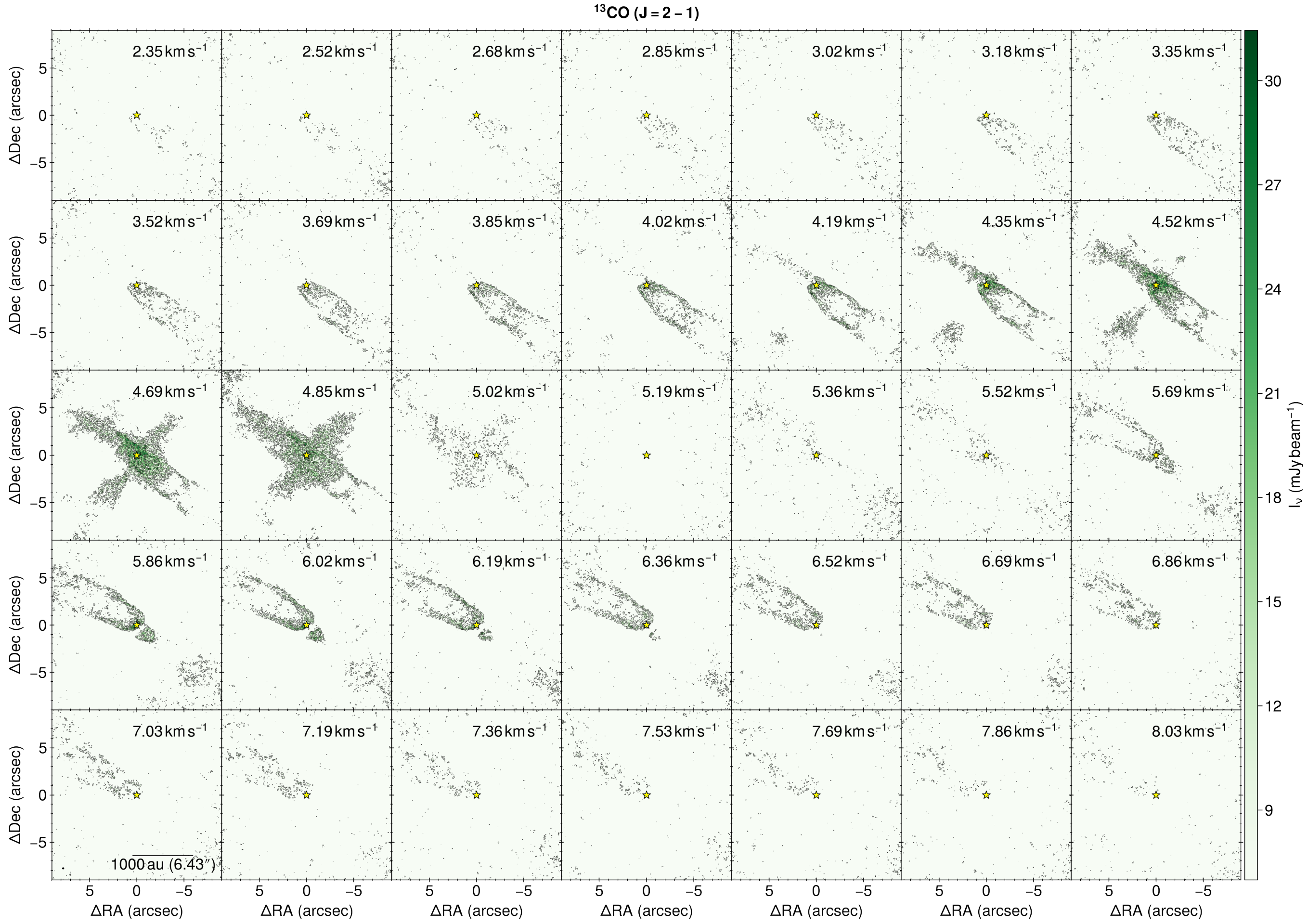

**Figure B.3.** Same as Figure B.1, but for $^{13}$CO ($J = 2 \rightarrow 1$). The channel map is centered at $5.19\,\mathrm{km\,s^{-1}}$

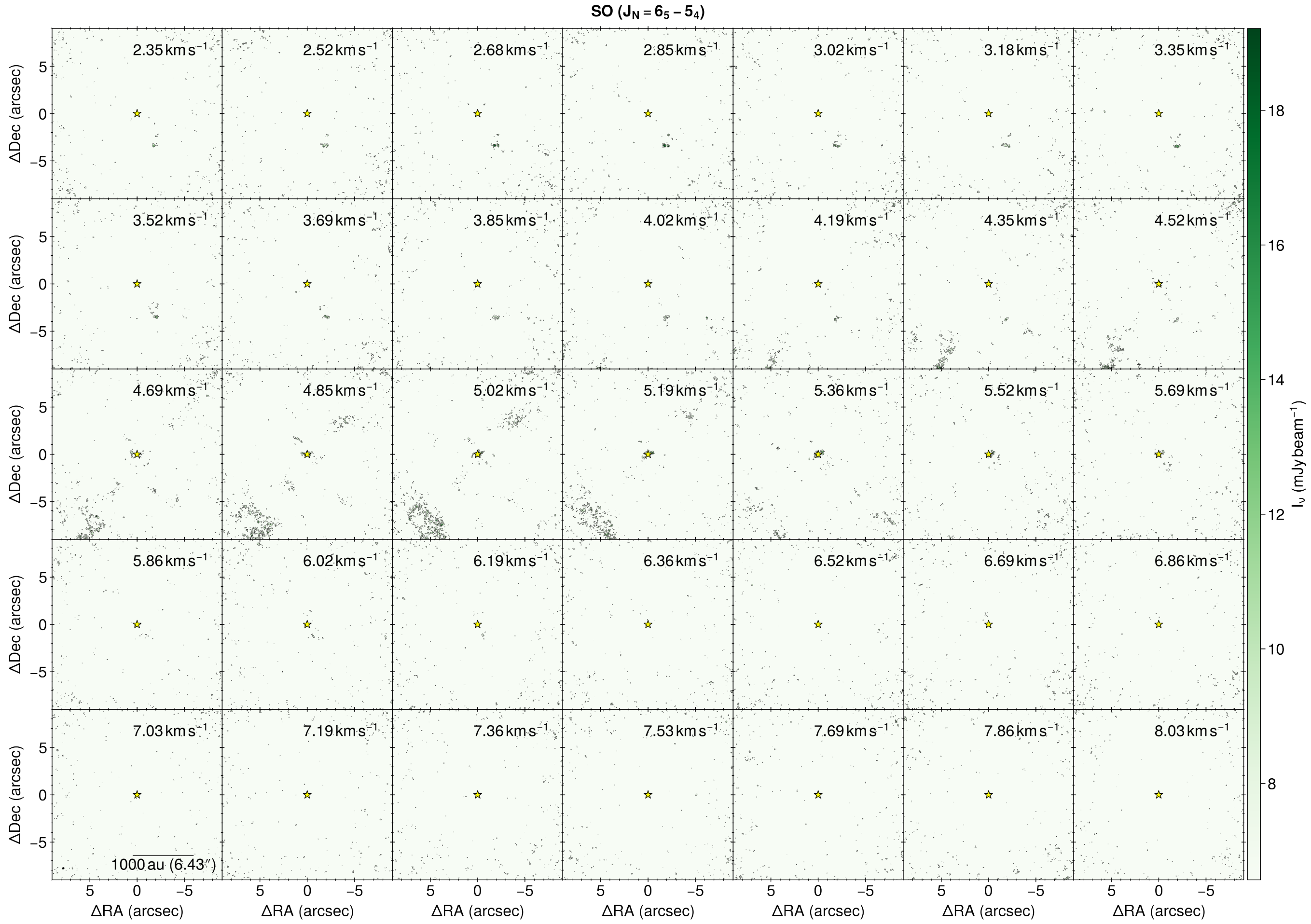

**Figure B.4.** Same as Figure B.1, but for SO ($J_N = 6_5 \to 5_4$). The channel map is centered at 5.19 km s$^{-1}$

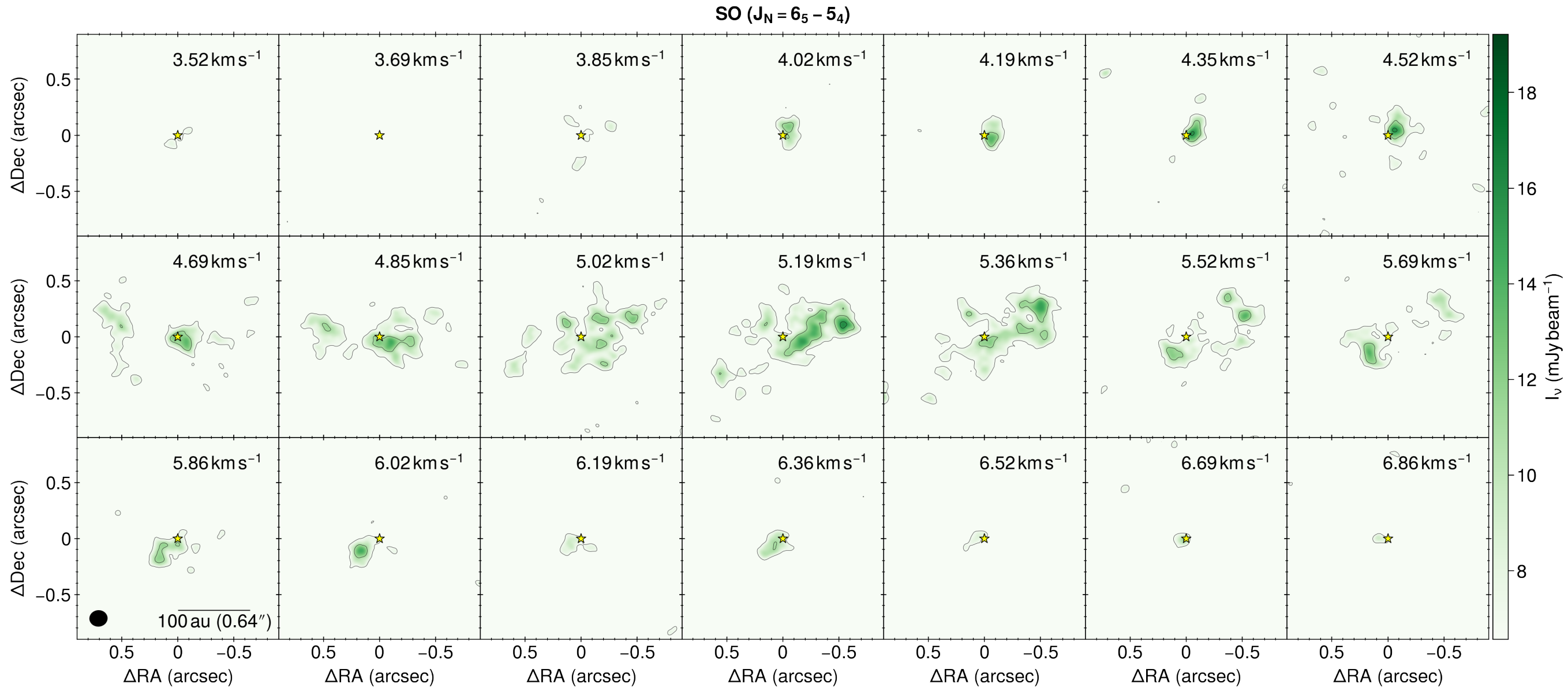

**Figure B.5.** Zoomed-in version of Figure B.4 for SO ($J_N = 6_5 \to 5_4$) showing velocities from 3.52 to 6.86 km s$^{-1}$. The channel map is centered at 5.19 km s$^{-1}$.

## C. SB-ONLY IMAGES OF MOLECULAR LINES

In Figures C.1 and C.2, we present the short baseline only (SB-only) integrated-intensity and intensity-weighted velocity maps for IRAS15398. The estimated rms of the $^{12}$CO, C$^{18}$O, $^{13}$CO, and SO integrated-intensity maps are 8.79, 1.93, 4.05 and 3.05 mJy beam$^{-1}$ km s$^{-1}$, respectively. Please see Section 3 for more description.

**Table C.1.** Summary of Representative SB-only Molecular Line Maps

| Molecular Line | Frequency (GHz) | Robust | Beamsize | Velocity Resolution (km s$^{-1}$) | Peak Intensity (mJy beam$^{-1}$) | RMS Noise (mJy beam$^{-1}$) |
|---|---|---|---|---|---|---|
| C$^{18}$O ($J = 2 \rightarrow 1$) | 219.56035 | +2.0 | $0''.337 \times 0''.301$ (+20.9°) | 0.167 | 46.2 | 2.98 |
| SO ($J_N = 6_5 \rightarrow 5_4$) | 219.94944 | +2.0 | $0''.338 \times 0''.301$ (+23.8°) | 0.167 | 35.82 | 3.58 |
| $^{13}$CO ($J = 2 \rightarrow 1$) | 220.39868 | +2.0 | $0''.338 \times 0''.295$ (+25.3°) | 0.167 | 105.5 | 4.09 |
| $^{12}$CO ($J = 2 \rightarrow 1$) | 230.53800 | +2.0 | $0''.324 \times 0''.288$ (+22.1°) | 0.635 | 216.6 | 1.98 |

Note—The values listed here were obtained from the maps using short baseline only (SB-only) observations. The peak intensity of the line maps were found by using a 2″circular aperture around the center position on the channel with the highest intensity. The rms noise of the line maps were found by using a 10″circular aperture in an emission-free area around the center position on a line-free channel near the beginning of the datacube.

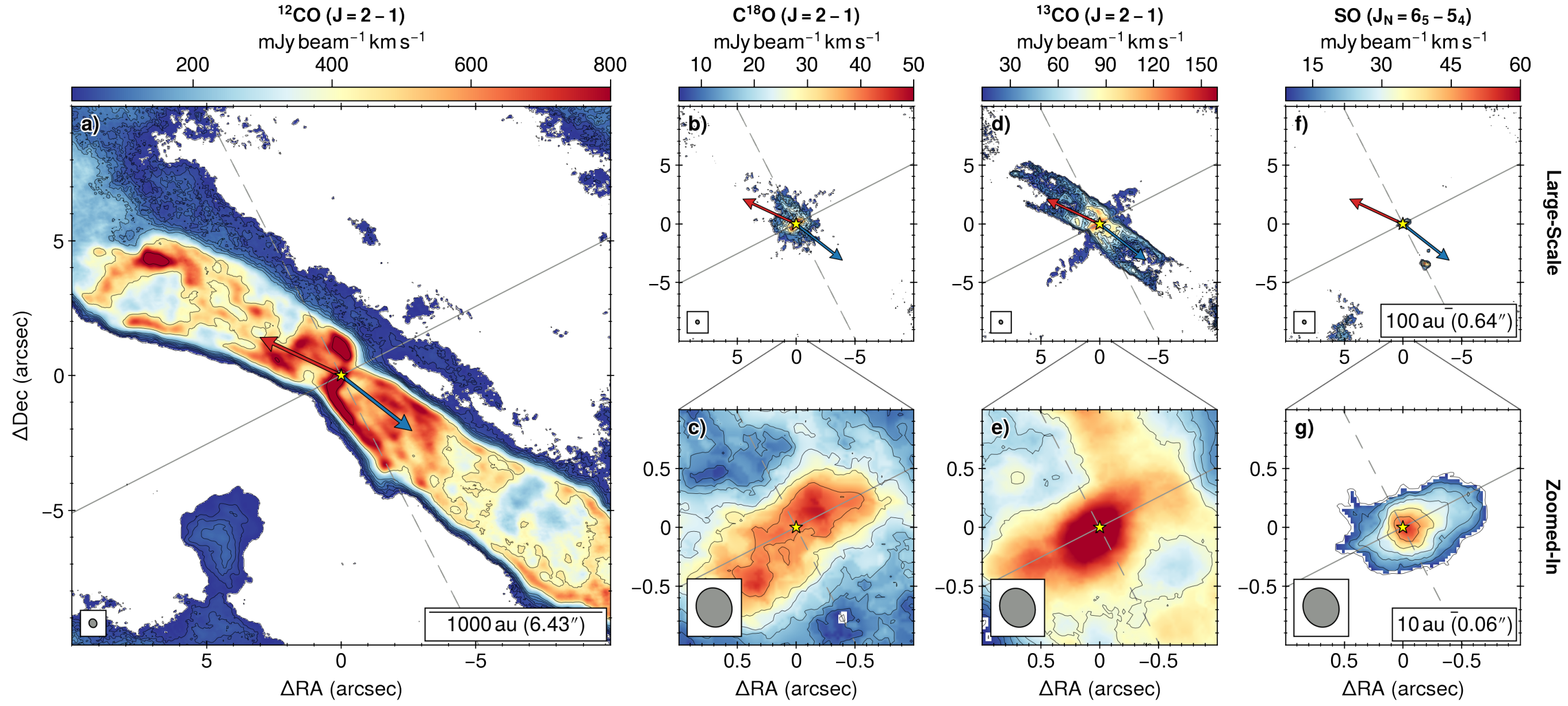

**Figure C.1.** Same as Figure 3, but for the SB-only maps of the representative molecular lines.

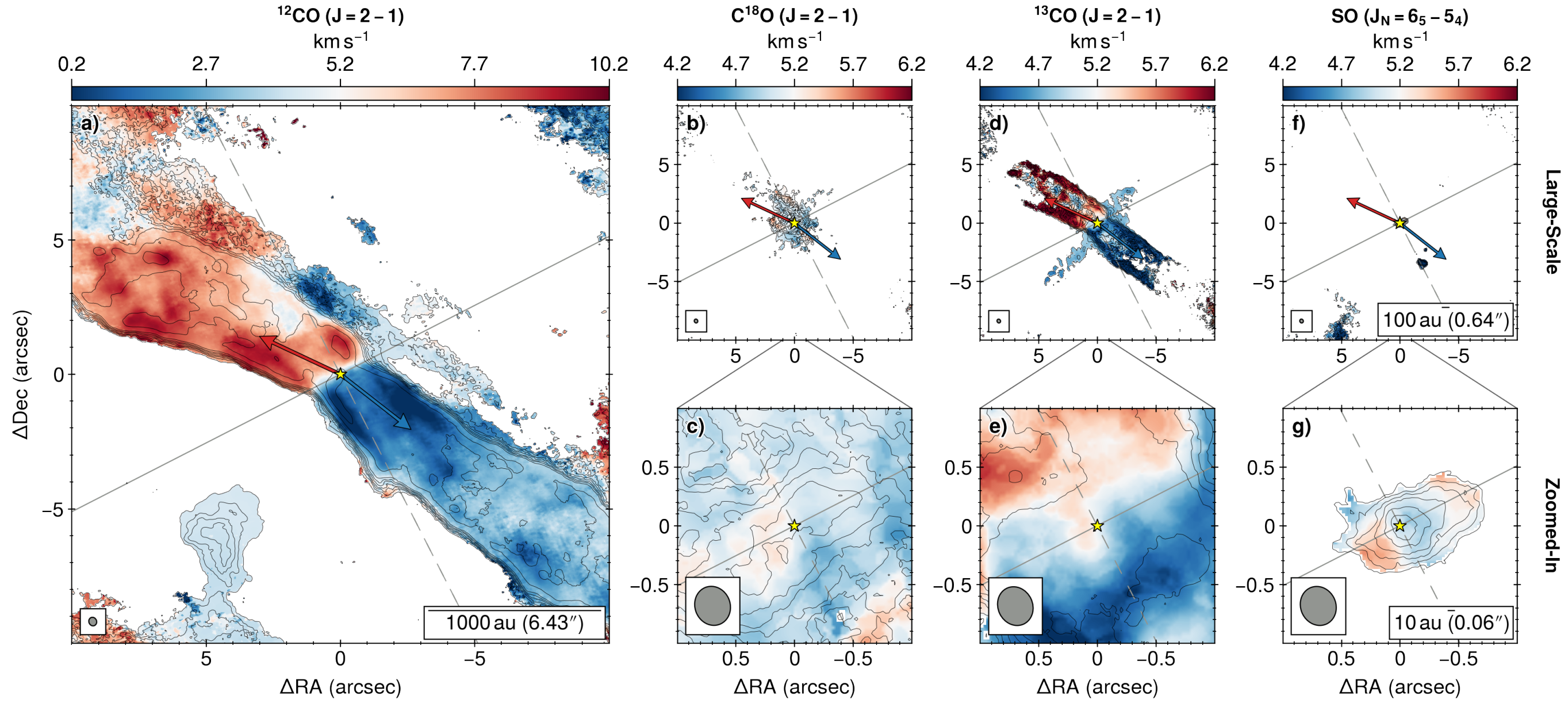

**Figure C.2.** Same as Figure 4, but for the SB-only maps of the representative molecular lines.